\documentclass{jfm}
\usepackage{graphicx}
\usepackage{newtxtext}
\usepackage{newtxmath}
\usepackage{natbib}
\usepackage{hyperref}
\hypersetup{
    colorlinks = true,
    urlcolor   = blue,
    citecolor  = black,
}
\usepackage{siunitx}
\usepackage[version=4]{mhchem}
\usepackage{color,soul}
\usepackage{mathtools}
\usepackage{textcomp,gensymb}
\usepackage{amsmath}
\usepackage[gen]{eurosym}
\DeclareSIUnit\pixel{px}
\DeclareSIUnit\minutes{min}
\DeclareSIUnit\euro{\euro{}}
\DeclareSIUnit\fps{fps}
\usepackage[scr=boondox,  
            cal=esstix]   
           {mathalpha}
\newcommand{\RomanNumeralCaps}[1]

\usepackage{graphicx}
\usepackage{subfig}
\usepackage{floatrow}

\shorttitle{Melting of floating ice cylinders}
\shortauthor{Edoardo Bellincioni, Detlef Lohse, Sander G. Huisman}

\title{Melting of floating ice cylinders in fresh and saline environments}

\author{Edoardo Bellincioni\aff{1}
  \corresp{\email{e.bellincioni@utwente.nl}},
  Detlef Lohse\aff{1,2},
 \and Sander G. Huisman\aff{1}}

\affiliation{\aff{1}Physics of Fluids Department and Max Planck Center for Complex Fluid Dynamics and J.M. Burgers Centre for Fluid Dynamics, University of Twente, P.O. Box 217, 7500AE Enschede, The Netherlands
\aff{2}Max Planck Institute for Dynamics and Self-Organisation, Am Faßberg 17, 37077, Göttingen, Germany}

\begin{document}

\maketitle

\begin{abstract}
Motivated by the need for a better understanding of the melting and stability of floating ice bodies, we experimentally investigated the melting of floating ice cylinders. Experiments were carried out in a tank, with ice cylinders with radii between \qty{5}{\centi\meter} and \qty{12}{\centi\meter}, floating horizontally with their axis perpendicular to gravity. The water in the tank was at room temperature, with salinities ranging from \qty{0}{\gram\per\liter} to \qty{35}{\gram\per\liter}.  These conditions correspond to Rayleigh numbers in the range 10$^5\lesssim$ Ra $\lesssim$ 10$^9$. The relative density and thus the floating behaviour was varied by employing ice made of H$_2$O--D$_2$O mixtures. In addition, we explored a two-layer stable stratification. We studied the morphological evolution of the cross-section of the cylinders and interpreted our observations in the context of their interaction with the convective flow. The cylinders only capsize in fresh water but not when the ambient is saline. This behaviour can be explained by the balance between the torques exerted by buoyancy and drag, which change as the cylinder melts and rotates. We modelled the oscillatory motion of the cylinders after a capsize as a damped non-linear oscillator. The downward plume of the ice cylinders follows the expected scalings for a line-source plume. The plume's Reynolds number scales with Rayleigh number in two regimes, namely Re $\propto$ Ra$^{1/2}$ for Ra $< \mathcal{O}(10^7)$ and Re $\propto$ Ra$^{1/3}$ for Ra $> \mathcal{O}(10^7)$, and the heat transfer (nondimensional as Nusselt number) scales as Nu $\propto$ Ra$^{1/3}$. Although the addition of salt substantially alters the solutal, thermal and momentum boundary layers, these scaling relations hold irrespectively of the initial size or the water salinity. While important differences exist between our experiments and real icebergs, our results can qualitatively be connected to natural phenomena occurring in fjords and around isolated icebergs, especially with regard to the melting and capsizing behaviour in stratified waters. 
\end{abstract}


\section{Introduction}\label{sec:01intro}
The lack of understanding of ice-sheet melting dynamics is the largest source of uncertainty in the projections of sea-level rise \citep{robel_marine_2019}\footnote{Throughout the introduction, the terms \textit{sea ice}, \textit{ice-sheet}, \textit{icebergs}, etc., are used as in the reference they correspond to. The physics and interaction mechanisms of each are different, and the terms are not interchangeable.}. The complexity of the phase-change physics and the uncertainties in the parametrisations (e.g. for albedo changes and for the interactions of the ice with ocean currents) implies that most of today's climate models poorly model the sea ice variations \citep{stroeve_arctic_2007,malyarenko_synthesis_2020}. The different contributors to sea level rise (among which, thermal expansion and ice sheets) are commonly assumed to be independent of each other, and this is deemed to be the reason for an underestimation of the uncertainty in sea-level projections \citep{le_bars_uncertainty_2018}. One of the processes that connect the ocean with ice structures is calving at the Greenland Ice Sheet (GrIS) and Antarctic Ice Sheet (AIS), where ice chunks up to hundreds of meters of length \citep{orlowski_chasing_2012} detach from the edge of a glacier, becoming icebergs. On top of being a major component in both the AIS and GrIS mass balance \citep{bigg_century_2014, cenedese_icebergs_2023}, icebergs and their freshwater fluxes have been proved to be a contributor to the freshwater balance in the Southern Ocean \citep{silva_contribution_2006}. An increase in the freshwater input from icebergs and meltwater discharge can have strong consequences on the global climate \citep{schloesser_antarctic_2019}. \par
Oceanographic surveys are precautionally kept far from icebergs, due to the unpredictability of capsizing events, thus limiting the data collection in close proximity of icebergs \citep{yankovsky_surface_2014}. Correspondingly, limited research has been conducted on freely floating melting objects. A historically significant branch of studies developed in the second half of the 20th century, when the interest in harvesting icebergs for freshwater motivated research on their melting during towing, see e.g. \citet{hult_antarctic_1973} and \citet{russell-head_melting_1980}. More recently, growing interest in iceberg research led to experimental and numerical investigations on the capsizing of icebergs. \citet{burton_laboratory_2012} and \citet{bonnet_modelling_2020} model natural icebergs as plastic rectangular cuboids, with well-defined aspect ratios, but excluded melting. Their research focuses on the capsizing events: how it is affected by the aspect ratio, how energy is released by a calving event, and how the capsizing triggers mixing and flow around the iceberg. \par
The essential physics of an isolated iceberg melting in the ocean is that of a Stefan problem \citep{rubinstein_stefan_1971}, where a cold ice mass lays on top of a warm, saline water reservoir. Three fields control the melt rate of the ice: temperature, salinity, and velocity of the fluid. First and foremost, the melting temperature of the water is a function of the salinity $T_\text{melt}=T_\text{melt}(S)$ and decreases with increasing salinity. Secondly, the Lewis number (Le $=\kappa_S/\kappa_T$, the ratio between saline diffusivity $\kappa_S$ and thermal diffusivity $\kappa_T$) controls the relative thickness of the thermal and boundary layers. Lastly, the density of water is itself a function of temperature and salinity, $\rho_\infty=\rho_\infty(T,S)$. The interactions between the melting and the water density are mutual, with the temperature and salinity fields that are affected by the melting ice, and the resulting density differences that set the water in motion, advecting water to the ice surface, whose temperature and salinity affect the melting itself. \par
Additional complexity arises if the free surface around the ice is considered, and the ice is unconstrained. In fact, in this configuration, the ice floats expose only a fraction of their volumes to air, and the floating behavior is controlled by the ratio between ice and water densities, $\Lambda=\rho_\text{ice}/\rho_\infty$. The lower heat transfer rate with air compared to water causes a significant difference in the melt rate between the submerged ice and the ice exposed to air. Furthermore, the ice's motion can be both in response to its own morphological evolution and to the liquid's motion, with the ice exposing different parts of itself to the liquid, depending on its orientation. \par 
Hitherto, most experimental lab-scale research has focused on the melting of submerged ice. The studies on boundary layers of submerged slabs are extensive \citep{josberger_laboratory_1981,carey_transport_1982}, with recent particular interest in the appearance of scallops, both in melting and dissolution problems \citep{davies_wykes_self-sculpting_2018,cohen_buoyancy-driven_2020}. At the same time, melting in symmetrical domains has received interest due to the relative simplicity of theoretical description of the heat transfer. For this reason, \citet{fukusako_melting_1992} and \citet{yamada_melting_1997} have investigated the melting of a fully-submerged fixed horizontal ice cylinder, for different ambient temperature and salinity conditions. They grew ice around a cooled pipe in a water bath and then let it melt, imaging the cross section with a SLR camera and visualising the flow using streakline photography. They concluded that the flow profile in the vicinity of the cylinder was laminar and bidirectional in the lower portion of the cylinder, and turbulent unidirectional in the upper portion of the cylinder, with the flow enhancing the melt rate. \par
Another line of research that has run in parallel to the aforementioned ones, is the study of plumes around a heated submerged object. The most relevant references here are the work of \citet{grafsronningen_piv_2011} and \citet{grafsronningen_simultaneous_2012,grafsronningen_large_2017}, who extensively investigated the plumes above a horizontal heated cylinder, both numerically and experimentally. They control the Rayleigh number by changing the temperature difference between the cylinder and the fluid, allowing them to explore Rayleigh numbers from \num{2.05e7} to \num{7.94e7}. \par
To the best of our knowledge, the only study that integrated the analysis of a floating melting object with a theoretical description of the heat transfer is by \citet{hosseini_experimental_2009}. The authors put an ice cylinder to float in a tank, and measured its cross-sectional width constraining it between vertical wires and measuring the distance between the wires. They find that the Rayleigh number (in their case, Ra $=\mathcal{O}(10^7)$) scales with the Nusselt number with a 1/3 power law, for experiments conducted in freshwater. We extend their results by analysing images of the cross-section, thus being able to deliver a more complete morphological analysis. \par
In several natural circumstances, icebergs (or ice chunks, of all sizes) melt in stratified waters. The most studied example is the one of Greenlandic fjords, where icebergs calve from a glacier and accumulate in the fjord, where they melt, releasing cold fresh water, which accumulates on top of of the warmer saline water. The draft of icebergs is various, and with it the fraction of mass immersed in each layer \citep{jackson_externally_2014,fitzmaurice_effect_2016}. A similar stratification mechanism applies to open-water icebergs, as described in existing studies on plumes around icebergs (e.g. \citet{helly_cooling_2011,yankovsky_surface_2014}). As described in \citet{yankovsky_surface_2014}, the meltwater from an iceberg accumulates around the iceberg itself, and the iceberg will find itself in a stably-stratified two-layer environment, where the less saline and thus lighter top layer has been produced by the accumulation of the iceberg's meltwater itself. \par
Recently, the problem of melting has been studied experimentally in several geometries and flow configurations. In \citet{fitzmaurice_nonlinear_2017} the authors used ice suspended in a flume to study the effect of oceanic currents on the melting, and highlighted the existence of two melting regimes, one with the melt plume attached and one with the melt plume detached. A similar setup was used in \citet{hester_aspect_2021}, where the authors compare the meltrate in the different faces of a cuboid iceberg, and show the dependency of the melt rate variations between the faces on the water velocity. The authors of \citet{meroni_nonlinear_2019} carried out experiments with ice melting in a rotating water tank, and found a relation between the Rossby number (Ro, nondimensionalised rotation speed) and the melt rate.  The authors of \citet{mccutchan_enhancement_2024} report the results for an ice ball that melts submerged in an enclosed tank, with varying water temperatures and turbulence conditions, and show, for quiescent fresh water, the appearance of a well-defined upward or downward plume, depending on the water temperature. In \citet{xu_buoyancy-driven_2024} the authors study the melting behaviour of a submerged ice cylinder whose axis is parallel to gravity, and report on the heat transfers and the scallops that appear on the surface of the cylinder itself. Lastly, \citet{waasdorp_melting_2024} elaborates on the problem of submerged melting of an object composed of a material (olive oil) which is immiscible in water, and derive theoretical descriptions for the boundary layer characteristics. The problem of \textit{floating} melting was explored by \citet{johnson_poster_2023} in a Gallery of Fluid Motion where they show that floating ice cylinders repeatedly capsize while melting. \par
In this paper we will focus on three aspects of the floating melting problem. \par
First, how the size of ice and the water's salinity affect the morphology of the ice. The melt characteristics of the ice are a direct consequence of the local heat transfer with the water, which is a function of salinity, temperature, and velocity fields. \par
Second, how the ice exchanges heat and momentum with the surroundings, in terms of (non-dimensional) heat transfers and plume characteristics. On these topics, we seek for scaling relations that summarise the physics.\par
Last, how stable the ice cylinders are against rotations around their axes. Both uneven melting and flow asymmetries can give rise to changes in the rotational stability of the object, which, in turn, causes oscillatory, rotational motion. \par 
This article is organised as follows. In Section \ref{sec:02methods} we provide our experimental methods. In Section \ref{sec:theory} we develop theoretical interpretations to the morphological evolution, convective heat transfers, plume, rotational stability and rotations. In Section \ref{sec:03results} we present and analyse our experimental results and put them in the theoretical framework. In Section \ref{sec:twolayer} we extend our findings to the more general two-layers stable stratifications. Lastly, in Section \ref{sec:04conclusions} conclusions and an outlook are provided. 

\section{Experimental apparatus and procedures}\label{sec:02methods}
\subsection{Experimental setup}
\begin{figure}
    \centering
    \includegraphics[width=.9\textwidth]{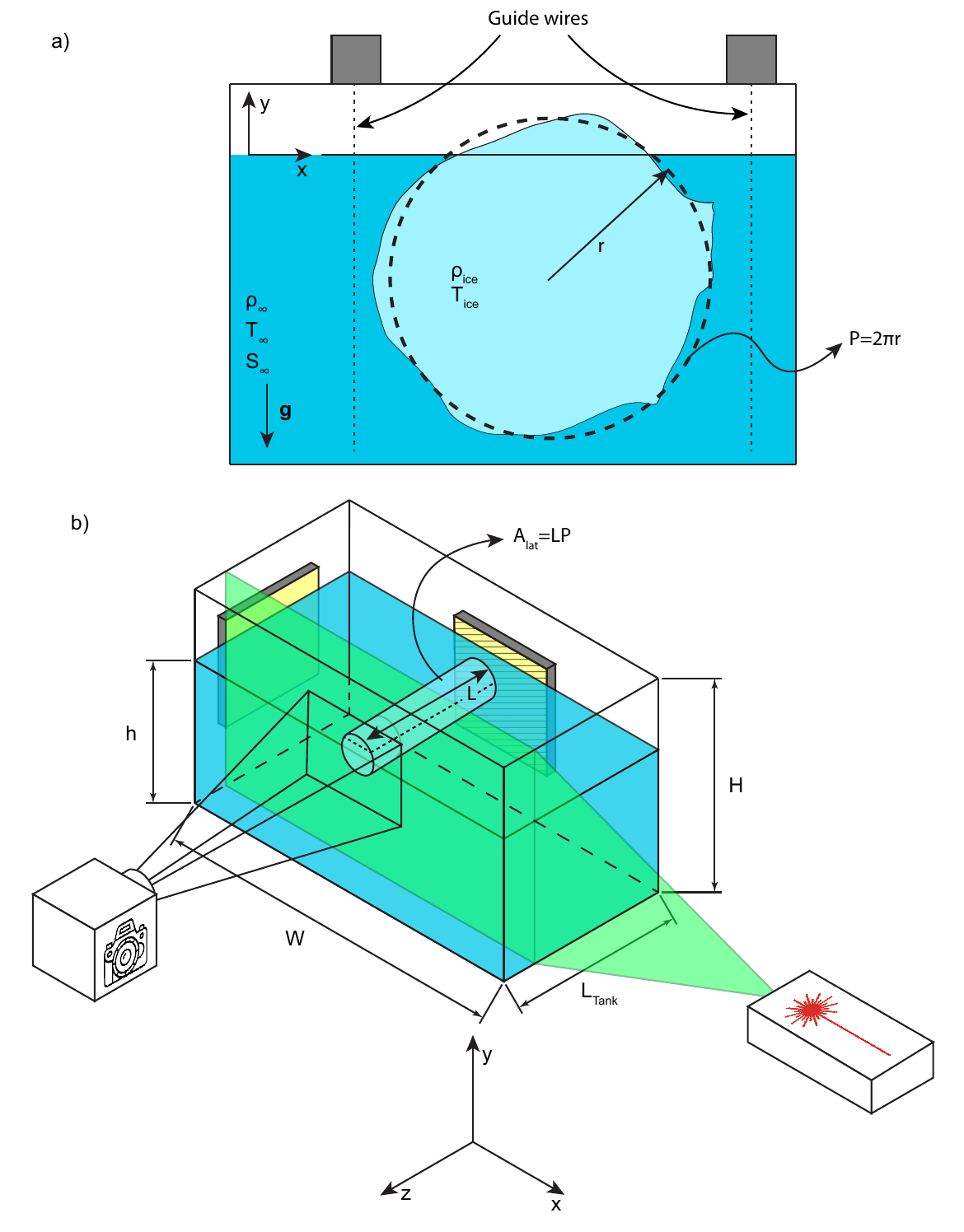}
    \caption{Sketches of the experimental setup. An ice cylinder (initial temperature $T_\text{ice}$, initial radius $R_i$) is left free to float on the surface of water (initial temperature $T_\infty$, initial salinity $S_\infty$, initial density $
    \rho_\infty=\rho(T,S)$) in an open tank. \textit{a)} Vertical nylon wires create a ``cage'' around the ice, with the ice being in contact with the wires only occasionally. The dashed circle sketches the equivalent radius of the ice, calculated as the radius of a disk that has the same area as the cross-section of the ice (with a radius $r$ and a perimeter $P=2\pi r$). \textit{b)} For contour tracing, we use background illumination from a linearly polarised LED light source from a panel at the back of the ice. A linear polarising filter, with the polarisation direction orthogonal to the light source, is screwed onto the objective of the camera. This configuration of polarisation only lets light through that passes through a birefringent material, which crystalline ice is. The side (non-polarised) panel shines light that illuminates the opaque parts of the ice cylinder, scattering non-polarised light, which is also captured by the sensor of the camera. For flow field measurements, we use planar PIV with the laser plane being vertical and perpendicular to the cylinder's axis. The length of the cylinders is indicated with $L$, and the lateral area as $A_\text{lat}=LP$.}
    \label{fig:setup}
\end{figure}
All the experiments were carried out in a  glass tank with outer dimensions W $\times$ L$_\text{tank}$ $\times$ H =\newline \qty{80}{\centi\meter} $\times$ \qty{40}{\centi\meter} $\times$ \qty{50}{\centi\meter} and \qty{8}{\milli\meter} thick walls. The tank was filled with water to a level of \qty{38.3}{\centi\meter}, for a total volume of \qty{115}{\liter}. The ice cylinders were put to float parallel to the short-side of the tank, centered with respect to the long side. To avoid the cylinder drifting around in the tank, vertical nylon wires bound the cylinder laterally see figure \ref{fig:setup}. The wires also prevent rotations around the vertical (as seen by \citet{dorbolo_rotation_2016, schellenberg_rotation_2023}), which would have been detrimental to our backlight illumination setup. {Note that we found that the thermal and dynamical interaction between the cylinder and the wire is not important for the melting dynamics, nor that the friction between the ice and the wires prevents possible rotational motion of the cylinders around their axes}. During the placement of the cylinders at the beginning of every experiment, care was taken to not excessively disturb the water. Nonetheless, standing and travelling waves were always present for the first few seconds of the experiments. To avoid any parallax problem in the imaging of the near sub-surface region for subsequent experiments, an equivalent amount of water was removed from the tank after the melting of every cylinder. Similarly, all the water was mixed  after every experiment, to avoid unwanted stratification or inhomogeneities in temperature or salinity. \par
We froze ice cylinders of different diameters: \qty{5}{\centi\meter}, \qty{8.1}{\centi\meter}, and \qty{12}{\centi\meter}, with a length of approximately \qty{30}{\centi\meter}, resulting in a total volume of \qty{0.5}{\liter}, \qty{1.5}{\liter}, and \qty{3.4}{\liter}, respectively. A summary of the explored parameters is shown in table \ref{tab:expTable}. The presence of tiny amount of bubbles in the ice (resulting from dissolved air in the water) seemed not to influence the melting process, in line with the findings of \citet{wengrove_melting_2023}. Despite this, we used de-ionised water (Milli-Q), which we put under vacuum with constant stirring for a sufficiently long time before freezing to remove the majority of the dissolved gas. The small (\qty{5}{\centi\meter} diameter) and large (\qty{12}{\centi\meter} diameter) cylinders were made using aluminium moulds, and the middle-sized (\qty{8.1}{\centi\meter} diameter) cylinders in PVC moulds. The freezer was kept at a temperature of \qty{-5}{\celsius}$\pm$\qty{0.1}{\kelvin}, to ensure a slow freezing process. When frozen, the cylinders were quickly removed from the freezer, and from the moulds, and put to thermalise in another freezer at \qty{-16}{\celsius}. After a day\footnote{The timescale for thermal diffusion in the ice ($r_\text{cyl}^2/\kappa_T$) is in the order of 1 hour}, they were removed and put to float in the tank. \par 
The bulk water temperature was measured at the beginning and end of every melting experiments with a K-type thermocouple (temperature sensitivity \qty{0.1}{\kelvin}). To measure the homogeneity, the temperature was probed in several locations of the tank, and the absolute variation was considered as the error. Note that the typical error was in the order of \qty{0.2}{\kelvin}, for a typical temperature of \qty{19}{\celsius}. \par
To vary the relative density between ice and water ($\Lambda = \rho_\text{ice}/\rho_\infty$) we modified our experiment in two ways. To explore ice that would float less, we performed experiments with deuterium oxide (D$_\text{2}$O, colloquially known as heavy water). {For those, a mixture of 81 wt\% D$_\text{2}$O and 19 wt\% H$_\text{2}$O was frozen.} The ice of this mixture closely matches the density of the surrounding water, $\frac{\rho_\infty-\rho_\text{ice}}{\rho_\infty}\approx1\%$, this made the ice float; just touching the surface, without breaking surface tension. \par
Similarly, to investigate ice melting in denser water, we added sodium chloride (NaCl) to the bulk water. This not only varies the ratio $\rho_\text{ice}/\rho_\infty$, but introduces added complexity as the fluid's density depends on both temperature and salinity ($\rho_\infty=\rho_\infty(T,S)$). \par
{To study the ice in absence of rotational motion, a 3D-printed plastic holder was frozen in one end of the cylinder.} The holder can slide along a vertical guide, allowing for vertical translation of the ice without rotation. The size and density of the holder made its impact on the buoyancy of the cylinder negligible. \par

\begin{table}
\begin{center}
{
\renewcommand{\arraystretch}{1.3}
\begin{tabular}{l|c|c|c|c}
                        & \multicolumn{3}{c|}{\ce{H2O}}                        & \ce{D2O}  \\
\cline{2-5}
D [\unit{\milli\meter}] & S = \qty{0}{\gram\per\liter} & S = \qty{10}{\gram\per\liter} & S = \qty{35}{\gram\per\liter} & S = \qty{0}{\gram\per\liter} \\ 
\cline{1-5}
\num{50}                & \num{1.6e7}                  & \num{6.5e07}                  & \num{1.9e+08}                 & \num{4.9e+08} \\
\num{81}                & \num{6.7e+07}                & \num{2.8e+08}                 & \num{8.0e+08}                 & n.a.          \\
\num{120}               & \num{2.2e+08}                & \num{9.0e+08}                 & \num{2.6e+09}                 & n.a.         
\end{tabular}
}
\caption{{Values of initial Rayleigh numbers (calculated according to equation \ref{eqn:Rayleigh}) for the explored cases of salinity and diameter, and the cases with deuterium oxide. Initial temperature of the water $T_\infty\approx$ \qty{20}{\celsius}, initial temperature of the ice $T_\text{ice}\approx$ \qty{-16}{\celsius}}. }
\label{tab:expTable}
\end{center}

\end{table}\par

\subsection{Measurements of shape}\label{subsec:shape}
In order to capture the shape and orientation evolution during the melting, a background illumination setup was arranged around the tank, as seen in figure \ref{fig:setup}. Two LED-light panels shine light from the side and back of the cylinder. The one in the back projects linearly-polarised light. {A DSLR camera (Nikon D850, \qty{45.7}{\mega\pixel}, pixel size \qty{4.35}{\micro\meter}) with a \qty{300}{\milli\meter} objective (Nikon AF-S Nikkor, resulting in a resolution of approximately \qty{50}{\micro\meter\per\pixel}) records the shadow of the cylinder.} A linear polarising filter is screwed on the objective of the camera, with the axis perpendicular to the one of the LED panel. This setup is most effective when high contrast is needed, but has not been used for all the experiments. The effect is based on the birefringent properties of \textit{clear} (crystalline) ice, which changes the polarisation direction of the incoming polarised light, thus letting some component through the two perpendicular polarising filters. On the other hand, the sharp index of refraction changes in \textit{opaque} ice (with lattice defects and microbubbles), leading to scattered light. Hence, the posterior (polarised) LED panel illuminates clear parts of the ice, while the lateral panel illuminates the opaque parts. The optical axis of the camera  was aligned to be perpendicular to the sides of the tank, and the cylinder aligned with the vertical wires to be in line with the optical axis. \par
Images were recorded at intervals of \qty{60}{\second} for the cylinders of \qty{5}{\centi\meter} and \qty{8.1}{\centi\meter} diameter, and \qty{180}{\second} for the cylinders of \qty{12}{\centi\meter} diameter, corresponding to the different melting rates. Contours were determined manually, with the aid of an assisted drawing tool (MATLAB \textit{drawassisted}). Each melting experiment produces 30 to 40 contours. The contours retain all the information about the 2-D shape of the ice cylinder, and the cross-sectional area and equivalent radius can be computed. Investigation on the minor length-wise variations of the morphology of the cylinder and end effects are out of scope for this work. \par
\subsection{Measurements of rotation dynamics} \label{sec:measRotDyn}
To investigate the rotational dynamics more precisely, videos of the ice rotations were recorded using the same lighting setup as described above, with a resolution of \numproduct{3840 x 2160} \unit{\pixel} at a framerate of \qty{30}{\fps}. Of these, every frame was analysed as described in subsection \ref{subsec:shape}. \par
\subsection{Fluid velocity measurements}
To investigate the convective flow arising from a melting cylinder we use a particle image velocimetry (PIV) setup, see figure \ref{fig:setup}. The beam of a  Litron LDY-300 Nd:YLF (\num{532} \unit{\nano\meter}) pulsed laser was widened into a vertical sheet with cylindrical optics (LaVision), and crossed the cylinder in a spanwise section, roughly at 1/3 of its length. The water was seeded with rhodamine B-coated PMMA particles with a diameter of \num{1}--\num{20} \unit{\micro\meter} from Dantec (PMMA-RhB-10). A Photron SA-X2 camera imaged the flow with a field of view of approximately \qty{15}{\centi\meter} $\times$ \qty{15}{\centi\meter} around the cylinder, at a framerate of \qty{50}{\hertz} and resolution of \qty{1}{\mega\pixel}. The synchronisation and triggering of the laser and camera were controlled through LaVision's DaVis software. The recording time was chosen to be \qty{40}{\second}, resulting in a total of \num{2000} frames. The camera buffer and data transmission allowed to make a recording roughly every \qty{5}{\minute}.
The PIV data were analysed using DaVis. Frames were undersampled at one every five (effective \num{10}\unit{\hertz}). The cross-correlation method was set to multi-pass (from \numproduct{32 x 32} \unit{\pixel} window to \num{16} $\times$ \num{16} \unit{\pixel} window with 50\% overlap), which corresponds to a final window size of \qty{1.22}{\milli\meter}. {Note that our PIV measurements were only done in non-stratified water. Furthermore, the regions of interest for our experiments are always outside the thermal BL around the melting ice, such that refractive index matching techniques are not required.}

\section{Theoretical background} \label{sec:theory}
\subsection{Morphological evolution: qualitative description}\label{sec:theo-morpho}
In this subsection we examine the evolution of the morphology and how it is affected by salinity in the background fluid. In the case of floating ice, in contact with air and (saline) water, the heat transfer mechanisms are radiation, conduction, and convection. We find radiation to be negligible\footnote{{$\sigma(T_\text{room}^4-T_\text{ice}^4)S_\text{exposed}\approx$ \qty{1}{\watt}, while convective flows, estimated from $\frac{\mathcal{L}\rho_\text{ice}V_\text{cyl}}{t_\text{melt}}$, are $\mathcal{O}(10^2$ \unit{\watt})}} compared to the other mechanisms. {Measured at \qty{20}{\celsius}, the thermal diffusivity of air is more than 100 times higher than that of water, and the air's thermal conductivity is roughly 25 times smaller than that of water. These differences imply that the conductive term in the heat balance is dominated by transfers with the water.} On top of that, the convective motions of the liquid triggered by buoyancy differences reduce the extent of the thermal boundary layers, further increasing the heat transfer rate. \par
In the freshwater case, both the meltwater and the cooled down surrounding water are denser than $\rho_\infty$, hence they sink, and while doing so they exchange heat with the cylinder, getting colder as they sink. {The local melt rate depends on the temperature gradient between the ice and the water, and is consequently decreases with depth.} \par

\begin{figure}
    \centering
    \includegraphics[width=\textwidth]{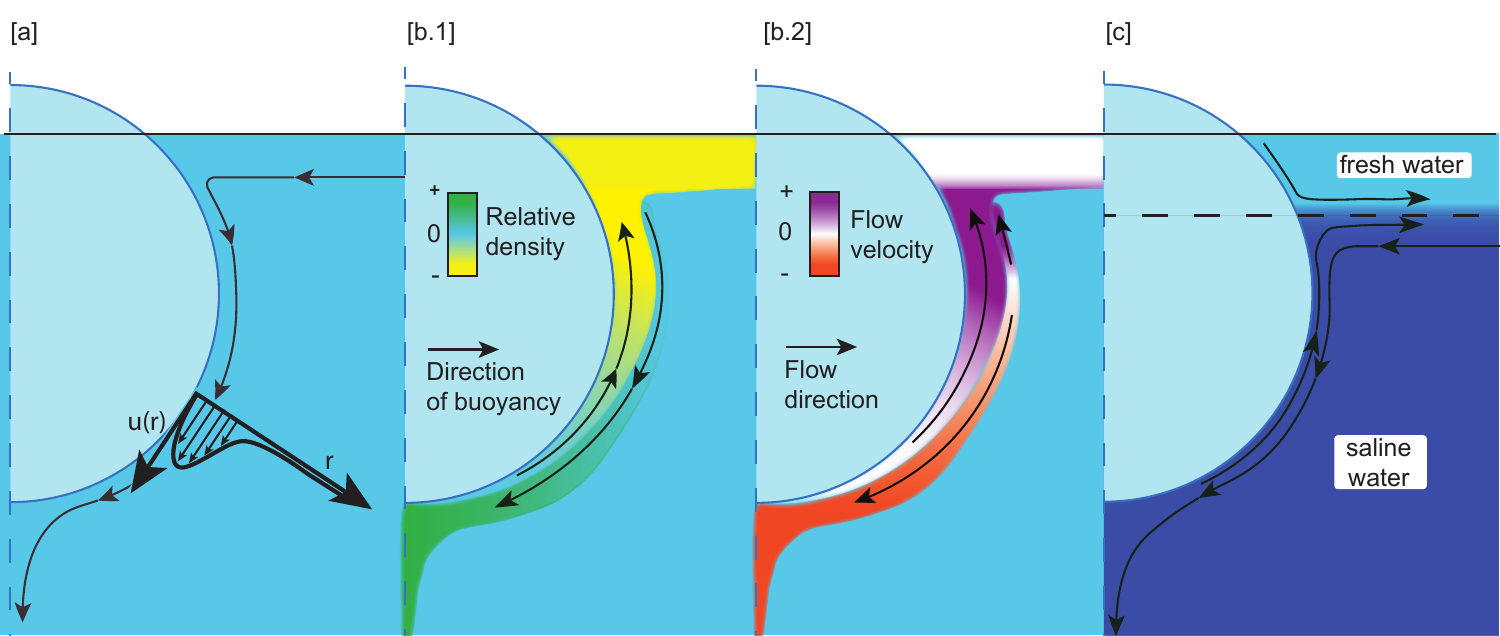}
    \caption{Sketch of the boundary layers around an ice cylinder melting in fresh, saline, and stratified water. In freshwater (panel a), a single direction of flow is developed. Both the cold meltwater and the cooled down surrounding water are denser, and sink, forming a plume below the cylinder. In panel b.1 we see how a sufficiently high salinity separates the flow in two directions, which carry two different density anomalies: the cold, fresh meltwater becomes lighter than the surrounding saline water, and rises, accumulating under the free surface; on the contrary, the cooled down, surrounding saline water becomes denser and sinks, forming a plume. The variation in intensity of the two flows along the cylinder implies that parts of the flow are bidirectional and other are unidirectional, see panel b.2. In the case of two-layered (stratified) water (panel c), the meltwater will accumulate across the density step, either being heavier than the top-layer water or lighter than the dense bottom-layer.}
    \label{fig:boundaryLayers}
\end{figure}
As for the saline case (for the salinities that we explored), the meltwater is less dense than $\rho_\infty$ (despite being much colder, the density is dominated by salinity), and tends to rise. The melting of the cylinder releases fresh meltwater, which further decreases the relative density of this innermost fluid. {At the same time, the much wider extension of the thermal boundary layer compared to the solutal boundary layer ($\delta_\text{thermal}/\delta_\text{solutal}\approx$ 20, as the Lewis number Le $=\kappa_T/\kappa_S \approx 400$ and $\delta_\text{thermal}/\delta_\text{solutal}\approx\sqrt{\text{Le}}$) implies that the bulk water decreases its temperature, and hence increases its density, see panel \textit{b.1} of figure \ref{fig:boundaryLayers}.} Just like in the freshwater case, the water continues to get cooled during its sinking motion, thus increasing its relative density. For most of the cylinder's latitudes, the flow in the vicinity of the ice surface is bidirectional, but there is a point where the positive buoyancy of the meltwater overcomes the negative buoyancy of the cooled saline water, and the flow becomes unidirectionally rising, see panel \textit{b.2} of figure \ref{fig:boundaryLayers}. Lastly, for the relative high salinities analysed here, the fresh meltwater accumulates below the surface, creating a cold, insulating layer which inhibits the melting.  \par 

\subsection{Heat transfer mechanisms}
A number of parameters play a role in the process of melting of a floating object, both for the fluid and for the melting solid. We use non-dimensional numbers to effectively compare experiments carried out under different configurations. The relevant non-dimensional numbers are the Rayleigh number, describing the intensity of the thermal driving, and the Nusselt number, describing the heat flux as compared to the conductive case. \par
{From the heat balance equation for a melting horizontal right prism, the Stefan condition reads }
\begin{equation}\label{eqn:StefanCondition}
    \frac{1}{P}\rho_\text{ice} \frac{\partial V}{\partial t}\mathcal{L} = \bigg\langle Lk_\text{ice}\frac{\partial T}{\partial n}\biggr\rvert_{-} - Lk_\text{water}\frac{\partial T}{\partial n}\biggr\rvert_{+}\bigg\rangle_P\,\,.
\end{equation}
{Here $V$ is the volume of the prism, $L$ its length, $P$ its cross-sectional perimeter, $\mathcal{L}$ the latent heat, $k=\kappa_T\rho c_p$ the thermal conductivity of the material, $c_p$ its specific heat, $\kappa_T$ its thermal diffusivity, $\frac{\partial}{\partial n}\big\rvert_{\pm}$ the spatial derivative in the direction normal to the ice surface, pointing into the water, evaluated at the interface, either in the ice (-) or in the water (+). The symbols $\langle\cdot\rangle_P$ indicate the average over the perimeter.} {The expression equates the heat flux from the water with the sum of those needed to melt the ice and that is diffused in the ice.} {The contributions of the ends are neglected (which we think is justified, given the aspect ratio of our cylinders).} \par
{Given the convective motions that arise in the vicinity of our cylinders, we can express the terms on the right hand side of equation \ref{eqn:StefanCondition} using an averaged convective heat transfer coefficient $h$}
\begin{align}
    \bigg\langle Lk_\text{ice}\frac{\partial T}{\partial n}\biggr\rvert_{-} - Lk_\text{water}\frac{\partial T}{\partial n}\biggr\rvert_{+}\bigg\rangle_P &= \frac{ h(T_\infty-T_\text{melt})A_\text{lat}}{P}, \\
    A_\text{lat}&=LP
\end{align}
{where $A_\text{lat}$ is the lateral surface of the ice. We assume here that the liquidus temperature $T_\text{melt}$ is equal to \qty{0}{\celsius}, hence $T_\infty-T_\text{melt}\approx T_\text{melt}$, which is a good approximation only for the salinity and temperature ranges explored in this work, and not for geophysical applications.} 
\par
Under the assumption that the melting occurs in the radial direction, the volume evolves as follows
\begin{equation*}
    \frac{\partial V}{\partial t} = L\frac{\partial A}{\partial t}
\end{equation*}
where $A$ is the cross-sectional area. Putting everything together, we obtain
\begin{equation}\label{eqn:heatTransfer}
    \rho_\text{ice} \frac{\partial A}{\partial t} \mathcal{L} = h T_\infty P.
\end{equation}
{If we define a typical length scale for a melting right prism of base area $A$ as the square root of $A$, and using equation \ref{eqn:heatTransfer} then the Nusselt number can be defined as }
\begin{equation}
    \text{Nu}=\frac{h\sqrt{A}}{k_\text{water}}= \frac{\rho_\text{ice} \sqrt{A} \frac{\partial A}{\partial t}\mathcal{L}}{k_\text{water} T_\infty P}\,\,\,.
\end{equation}
Under these assumptions for the length scales, the expression for the Rayleigh number is
\begin{equation}\label{eqn:Rayleigh}
    \text{Ra} = \frac{g\Delta\rho\left(\sqrt{A}\right)^3}{\kappa_T\nu\bar{\rho}}
\end{equation}
with $\Delta\rho/\bar{\rho}$ the density difference between the meltwater and the far-field water, normalised by the average between the two densities $\Bar{\rho}$, $g$ the  gravitational acceleration, and $\nu$ the kinematic viscosity of water.\par
{The expressions for Nu and Ra derived so far hold for any right prism. However, in the case of a right prism with a polygonal base with sharp angles, corrections due to the Gibbs--Thomson effect could be needed. }\par
The densities of the ambient water are calculated with the formula provided in \citet{millero_density_2009}, which accounts for the temperature and salinity of the water. For the heavy water cases, we estimated the density of our frozen mixture of H$_2$O and D$_2$O as a linear combination of the two densities. However, we could not find a valid measure of the density of heavy water ice for temperatures below \qty{0}{\celsius}, hence we assumed the ice to be at \qty{0}{\celsius}\footnote{{Our best estimate is that the temperature dependency of the density of heavy water D$_2$O behaves like that of ``normal'' water H$_2$O, with a $\approx$0.3\% difference between the density at \qty{-16}{\celsius} as compared to that at \qty{0}{\celsius}.}}. \par 
\subsection{Line-source plumes}
As sketched in figure \ref{fig:boundaryLayers}, density differences generate downward flows, both in the fresh and saline water experiments. {Such flows interact with the quiescent water in the tank, and are expected to show similar dynamics as line-source plumes. Hereafter we adapt classic plume theory (\citet{straneo_dynamics_2015}, and references therein) for fully-developed, turbulent line-source plumes, to obtain simple expressions for the plume width and velocity, as a function of depth. }\par
{The theory assumes that a localised, 2D source of mass and negative buoyancy generates a region of downward motion in a fluid, where the width $b(y)$ and velocity $w(y)$ can be calculated from the entrainment coefficient $\alpha$ and the source's mass and buoyancy fluxes ($Q$ and $B$, respectively), both per unit length}\footnote{{The length of our cylinders can be assumed constant as the axial melt rate is much slower than both the typical plume velocity and the radial melt rate. }}. {Under the assumption that the ambient water density is homogeneous, the plume buoyancy flux $B$ is predicted to be constant with depth. } {For the plume width $b$, the theory predicts a linear relation with depth $y$, with a slope equal to the entrainment coefficient $\alpha$, thus $b=\alpha y$. In our case, this expression will have with a virtual origin correction $\Tilde{y}$ ($\Tilde{y}$ such that $w(\Tilde{y})=0$ \citep{cenedese_entrainment_2014}) from our cylinders having a non-zero horizontal extent. } {For the plume velocity, the theory predicts the expression}
\begin{equation}
    \label{eqn:plume}
    w = \left(\frac{B}{2\alpha} \right)^{\frac{1}{3}}
\end{equation}
{that is constant in depth under the assumptions stated before.} {Additionally, given the definition and theoretical prediction for the reduced gravity in the plume $g'(y)$ }
\begin{equation*}
    g'(y):=g\frac{\rho_p(y)-\rho_\infty}{\rho_0}=\left(\frac{B}{2\alpha}\right)^{\frac{2}{3}}\frac{1}{y}
\end{equation*}
{here $\rho_p(y)$ is the density in the plume, $\rho_\infty$ the ambient density, and $\rho_0$ a reference density (taken here as the mean of $\rho_p$ and $\rho_\infty$), the plume velocity can then be re-written as }
\begin{equation}\label{eqn:plumeVelocity}
    w = \sqrt{yg'(y)}=\sqrt{yg\frac{\rho_p(y)-\rho_\infty}{\rho_0}} \,\, ,
\end{equation}
{highlighting the relation between the velocity and the plume density.}\par
To compare the velocity of the plumes generated by experiments with different Rayleigh number (equation \ref{eqn:Rayleigh}), we define the Reynolds number as
\begin{equation}\label{eqn:Reynolds}
    \text{Re} = \frac{w\sqrt{A}}{\nu}
\end{equation}
with $w$ the typical plume velocity, and $A$ the cross-sectional area. Note that the Rayleigh number, the driving parameter of the problem, is based on the cylinder, while the Reynolds number, the response parameter of the problem, is based on the velocity of the plume.
\subsection{Conditions for stability of cylinders}\label{sec:stabIndex}
At melting temperature, the density of liquid water is 9\% higher than that of crystalline ice, so whenever left free to float, ice exposes about 9\% of its volume to air (the literal tip of the iceberg). The higher meltrate of the immersed ice compared to the ice exposed to air (see subsection \ref{sec:theo-morpho}), makes the ice bottom-heavy (with gravity effectively acting upwards). In the case of an ice cylinder, the stability dynamics will manifest itself through rotations about the cylinder's axis of symmetry. {The cylinder (or any irregular right prismatic shape) will rotate around its axis until a stable equilibrium is found. } \par
\begin{figure}
    \centering
    \includegraphics[width=.75\textwidth]{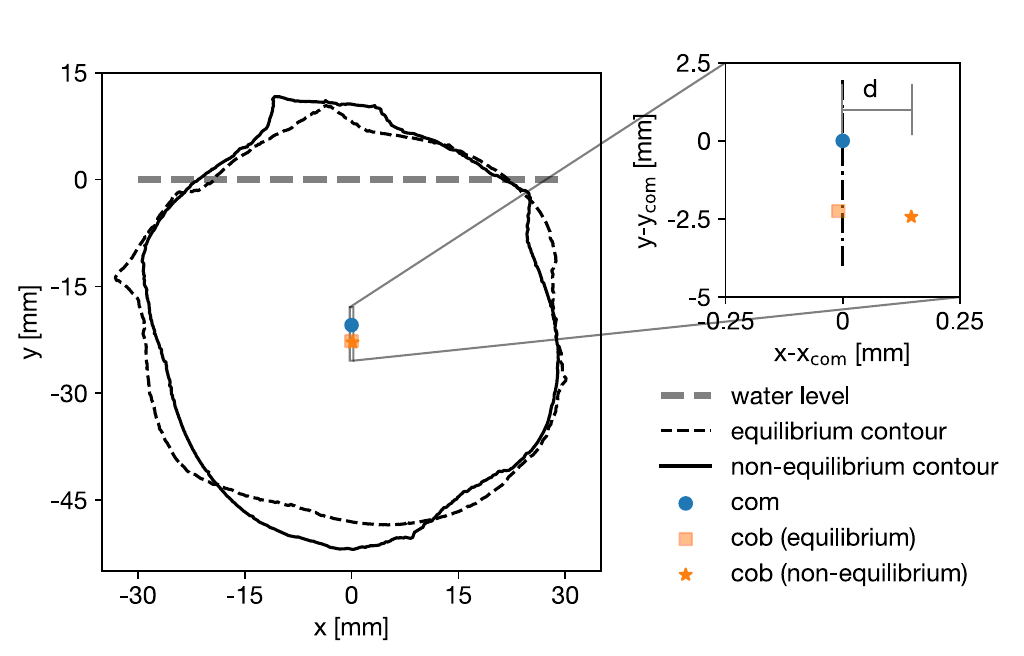}
    \caption{Gravitational and buoyancy effects on a floating melting cylinder (D$_\text{equivalent}$ = \qty{5.9}{\centi\meter}, D$_\text{initial}$ = \qty{8.1}{\centi\meter}, T$_\infty$ = \qty{19.3}{\celsius}$\pm$\qty{0.2}{\kelvin}, S = \qty{0.0}{\gram\per\liter}, Ra$_\text{initial}$ $\approx$ \num{4e7}, t $\approx$ \qty{10}{\minutes}). The dash-dotted horizontal line is the water level. The blue and orange dots (visible also in the insets) indicate the center of mass (com) and center of buoyancy (cob), respectively. In the inset, the distance $d$ is the horizontal distance between the two centers. The inset shows that in an equilibrium position the com and cob are vertically aligned (up to experimental accuracy), but the cob gets displaced if the shape is rotated around the com. Note that the horizontal scale is 10x stretched to highlight the precision of the experiment and to visualize the horizontal distances easier. }
    \label{fig:forcesRotation}
\end{figure}
\begin{figure}
    \centering
    \includegraphics[width=.8\textwidth]{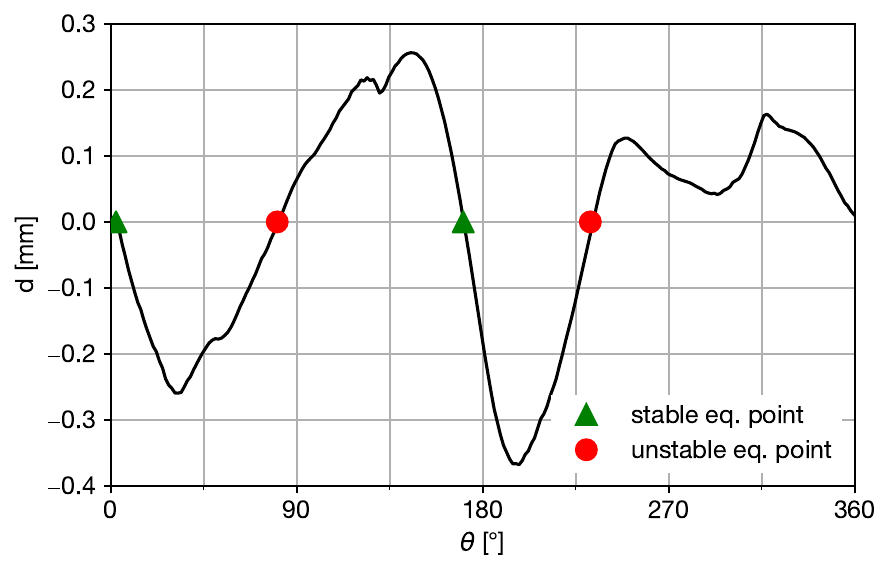}
    \caption{Horizontal displacement d of com and cob (d $=\text{com}-\text{cob}$, along the horizontal axis) as a function of the rotation angle $\theta$ for the shape in figure \ref{fig:forcesRotation}. The points where the curve crosses zero are equilibrium points. If the first derivative of the curve is positive (negative) in an equilibrium point, the buoyancy-induced torque is acting in the same (opposite) direction of the angle perturbation, hence the equilibrium point is unstable (stable). Stable equilibrium points are indicated with green triangles, while unstable with red circles. }\label{fig:stabilityDef}
\end{figure}

{Any floating object is subject to gravity and buoyancy, which can be considered as acting on two ideal points, the center of mass (com) and center of buoyancy (cob). If the object is in an equilibrium position, the com and cob are vertically aligned. }When a torque is applied, the induced rotation horizontally displaces the center of buoyancy, which generates a torque that can stabilise or destabilise the object, see the sketch in figure \ref{fig:forcesRotation}. For rotations happening around the center of mass, the arm of gravity's torque is zero. {We call $d$ the horizontal displacement between com and cob ($d=x_\text{com}-x_\text{cob}$), see figures \ref{fig:forcesRotation} and \ref{fig:stabilityDef}. Given our signs convention, in a stable equilibrium, an increase in the rotation angle will produce a negative \textit{d}, with the corresponding buoyancy torque that will act in the opposite direction as the angle increase, thus stabilizing the shape. On the contrary, in an unstable equilibrium, an increase in the rotation angle will produce a positive \textit{d}, with the buoyancy torque acting in the same direction as the angle increase, thus destabilizing the shape. } \par
It is important to notice that in order to destabilize a shape, the angular displacement needs to reach at least the closest unstable equilibrium point, otherwise the restoring force will push back the shape to where it was. \par
Two mechanisms control the stability of the ice against rotations: the magnitude of the displacement \textit{d} (which is responsible for the magnitude of the torque) and the angular distance to the closest unstable equilibrium point. {This argument is expanded in Appendix \ref{app:b}.} \par
With the aim of finding a quantitative measure of the stability which accounts for both these processes, we defined the stability index (SI) as the angular distance to the closest unstable equilibrium point multiplied by the maximum horizontal distance between the com and the cob throughout a full rotation of the shape. 

\begin{equation}
    {\text{SI}(\theta) = (\theta_\text{closest unstable}-\theta)\, \max_\theta(d(\theta))\, g\,\rho_\text{ice}\, A}
\end{equation}

with $A$ the cross-sectional area. {The choice of using $\max_\theta(d(\theta))$ arises from the fact that in both a stable and an unstable equilibrium point, the com and cob are vertically aligned, so $d(\theta)=0$, and this would contrast with one being more stable than the other. }\par
{Note that we do not nondimensionalize the stability index  for two reasons: first, applying a uniform scaling to the cylinders  changes the magnitude of the forces and torques involved; and second, the stability that the cylinders reach just before a rotation cannot be assumed universal, as the magnitude of the (random) perturbations that trigger a rotation is not. Hence, to} maintain physically-meaningful dimensions, we multiply this times the buoyancy force per unit length. Consequently, our stability index has the dimensions of a torque per unit length ($[N]$).\par 

\subsection{Rotation dynamics}
We now develop a dynamical model to describe the rotation dynamics of a capsizing event. We modelled the rotational motion as a damped, rotating non-linear oscillator. Newton's second law for a rotating object states 
\begin{equation}
    {I \Ddot{\theta} \hat{\boldsymbol{e}}_z= \sum \boldsymbol{\tau} = \sum \boldsymbol{a} \times \boldsymbol{F},}
\end{equation}
with $I$ being the moment of inertia of the object, and $\sum \boldsymbol{\tau}$ the sum of the torques exerted on the object ({$\boldsymbol{a}$ being the arm of each force $\boldsymbol{F}$}). It is assumed that the mass of the displaced fluid due to the rotational motion is negligible compared to the mass of the cylinder, hence no added mass is considered in our model. Consequently, under the assumption of cylindrical cross section the moment of inertia is simply $I=\frac{1}{2}mr^2$, with $r$ the equivalent radius. The deviation from the actual moment of inertia (calculated from the cross section) is of the order of 1\%. The cylinder is rotating around its center of mass (com) under the effect of the buoyancy force $\boldsymbol{F}_B$, that acts at a distance $d$ from the com (note that $d = d(\theta)$, see figure \ref{fig:forcesRotation}). Therefore, $\boldsymbol{F}_B = -\rho_\infty \boldsymbol{g} V_\text{sub}$, with $V_\text{sub}=(\rho_\text{ice}/\rho_\infty)V_\text{cyl}$. The corresponding torque will be $\boldsymbol{\tau}_B = \rho_\infty V_\text{sub} \boldsymbol{d}(\theta) \times \boldsymbol{g} $. {Lastly, the cylinder is subject to inertial drag, and as such to a drag force that acts with an arm equal to the radius of the cylinder, with a resulting torque of $\boldsymbol{\tau}_D = -\frac{1}{2}A_\text{lat}\rho_\infty C_D r^2\dot{\theta}|\dot{\theta}| \boldsymbol{r} \times \hat{\boldsymbol{e}}_\theta $, with $A_\text{lat}$ being the lateral surface of the cylinder and $C_D$ a drag coefficient. }\par
The resulting ODE for the damped nonlinear oscillator thus reads
\begin{equation}
    {
    \frac{1}{2}\:\Lambda\:V_\text{cyl}\:r^2\:\Ddot{\theta}=-g\:d(\theta)\:\Lambda\:V_\text{cyl}-\frac{1}{2}\:A_\text{lat}\:C_D\:r^3\:\dot{\theta}\:|\dot{\theta}|\,\,, 
    }
    \label{eqn:motionRotation} 
\end{equation}
\begin{equation}
    \Ddot{\theta} = -\gamma\:d(\theta) - \frac{2\:C_D}{\Lambda} \:\dot{\theta}\:|\dot{\theta}| \label{eqn:fittedRotation}
\end{equation}
with the density ratio $\Lambda = \frac{\rho_\text{ice}}{\rho_\infty}$ and $\gamma = \frac{2g}{r^2}$.\par

\section{Experimental results}\label{sec:03results}
\subsection{Morphological evolution}\label{sec:03.1shape}

\begin{figure}
    \centering
    \includegraphics[width=0.7\textwidth]{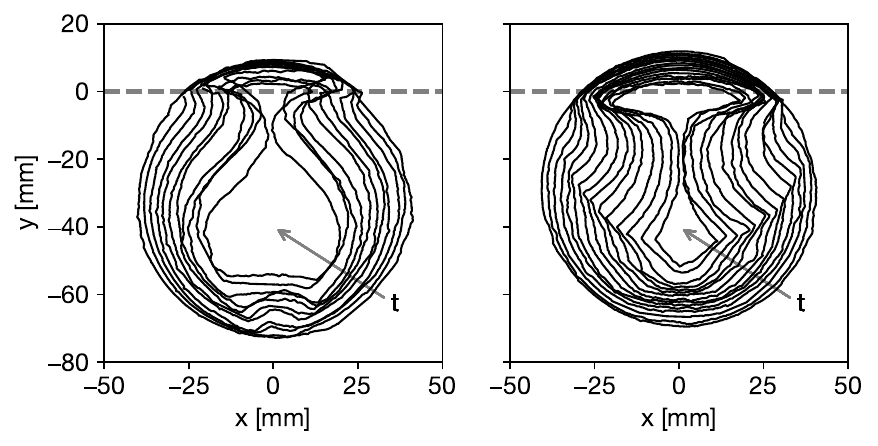}
    \caption{Shape evolution of (\textit{left}) a cylinder in fresh water whose rotations where prevented (D$_\text{initial}$ = \qty{8.1}{\centi\meter}, Ra$_\text{initial}$ $\approx$ \num{4.8e7}, T$_\infty$ = \qty{20.3}{\celsius}$\pm$\qty{0.2}{\kelvin}, S = \qty{0.0}{\gram\per\liter}) and (\textit{right}) a freely-floating cylinder (D$_\text{initial}$ = \qty{8.1}{\centi\meter}, Ra$_\text{initial}$ $\approx$ \num{1.7e8}) melting in saline water (T$_\infty$ = \qty{18.5}{\celsius}$\pm$\qty{0.2}{\kelvin}, S = \qty{10.0}{\gram\per\liter}). Contours are spaced by \qty{90}{\second}. Note that the two cylinders are at a different height as a result of the different $\rho_\infty$.}\label{fig:additionalShapes}
\end{figure}

First we will compare the morphological dynamics for the case of saline water to that of fresh water. We observed that ice cylinders melting in fresh water repeatedly capsized, regardless of their initial size, while the ones melting in saline water did not. Because of these different rotational dynamics, we performed an experiment constraining a cylinder in fresh water such that rotations would be prevented, but vertical motion would be allowed. Figure \ref{fig:additionalShapes} shows the evolution for both cases, and we readily see major differences in the morphological dynamics. Note here that in the freshwater case, the meltwater sinks, while in the saline water case, the meltwater floats. The constrained freshwater case shows an enhanced melt rate in the subsurface region and in the south pole; the saline water cylinder shows reduced melt rate in the subsurface region, and the appearance of a marked minimum of melt rate (showing as a protrusion) that migrates towards the south with time. These morphological features can be explained with arguments based on the boundary layer characteristics in the different cases, see figure \ref{fig:boundaryLayers}. \par

The feature that appears at the south pole of the freshwater cylinder is an effect of the recirculation zone: for high enough Rayleigh numbers, the sinking flow detaches from the cylinder, creating a turbulent recirculation zone below the object that enhances mixing and heat transfer; in later times (with lower Rayleigh numbers), the flow laminarises, but the footprint of the recirculation zone remains visible. This does not appear in saline conditions because the effective Rayleigh number is decreased as a result of the bidirectional flow, as depicted in panel (\textit{b.2}) of figure \ref{fig:boundaryLayers}. \par
{For the case of saline water, the minimum of melt rate is the morphological feature that arises because of the flow inversion identified in panel (\textit{b.2}) of figure \ref{fig:boundaryLayers} and also noted by \citet{yamada_melting_1997}}. Lastly, the reduced subsurface melt rate is a consequence of the accumulation of cold meltwater at the free surface. \par
\begin{figure}
    \centering
    \includegraphics[width=1\textwidth]{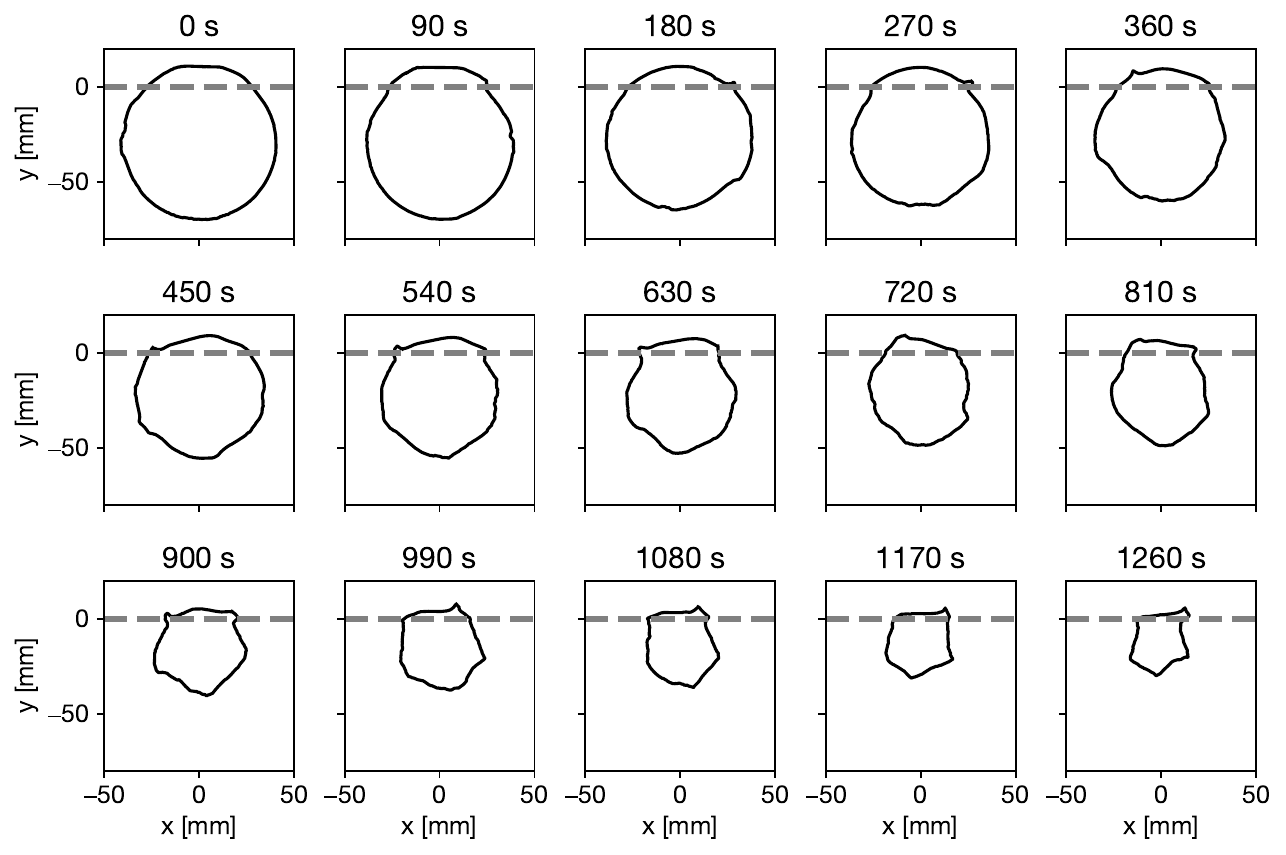}
    \caption{Shape evolution of an ice cylinder (D$_\text{initial}$ = \qty{8.1}{\centi\meter}, Ra$_\text{initial}$ $\approx$ \num{5.1e7}) melting in fresh water (T$_\infty$ = \qty{20.9}{\celsius}$\pm$\qty{0.2}{\kelvin}, S = \qty{0.0}{\gram\per\liter}). Each panel is labelled with the time from the beginning of the experiment. {The variation of the shape of the cross section along the cylinder is negligible.}}
    \label{fig:freshWater}
\end{figure}

We now come  to the more general, unconstrained ice melting in freshwater. In this case, the cylinder capsizes. The shape evolution is shown in figure \ref{fig:freshWater}. The different conductive heat transfers (air and water) lead to a marked asymmetry in melt rate, with the top of the ice melting significantly slower than the bottom. {This sculpting processes can create a horizontal asymmetry in the distribution of mass above and below the water surface, that can result in an imbalance between buoyancy and gravity, thus making the ice unstable against rotations around its axis.} The continued erosion makes the ice repeatedly capsize. The interaction between the flow and the ice surface sharpens the initial cylinder to a roughly polygonal prismatic volume. \par

\subsection{Convective heat transfers} \label{sec:03.4convection}
\begin{figure}
    \centering
    \includegraphics[width=\textwidth]{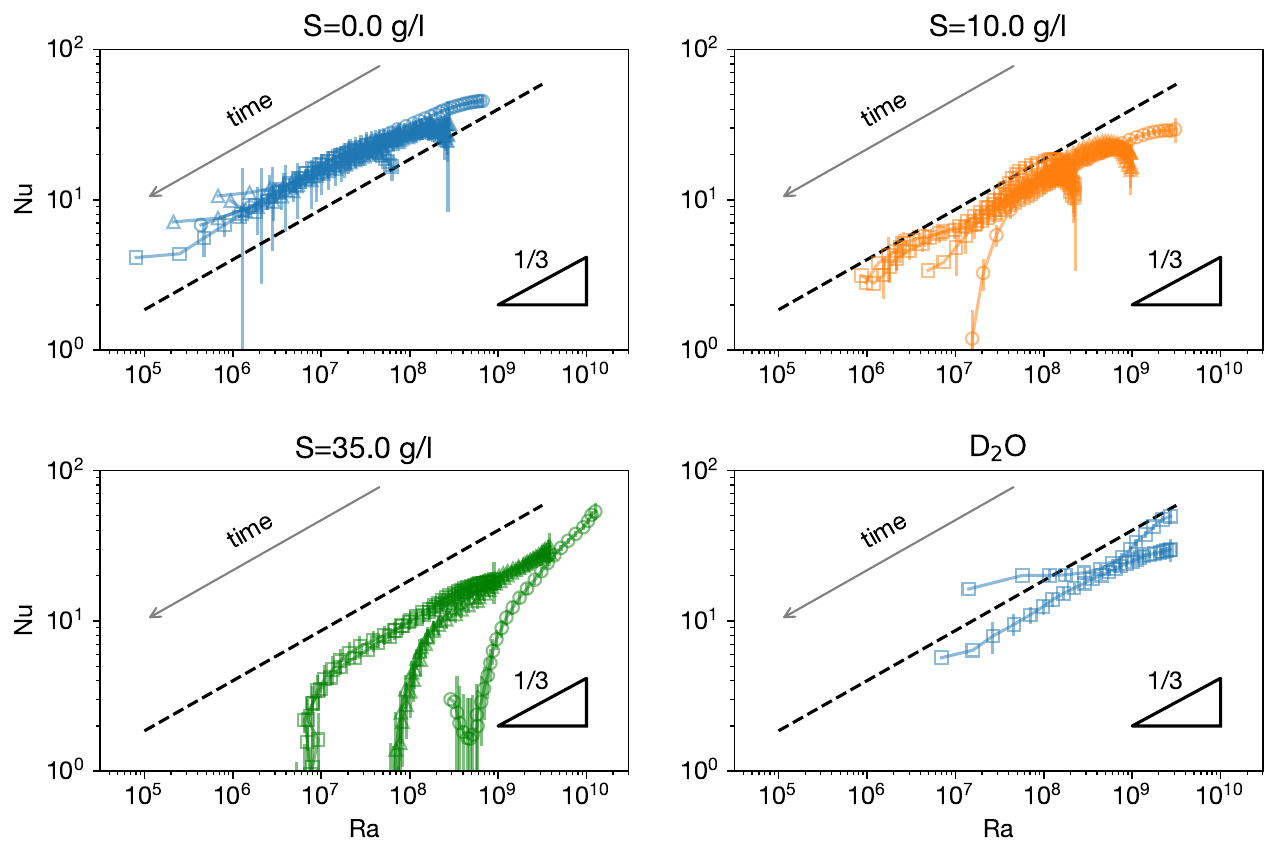}
    \caption{Scaling of the Nusselt number as a function of the Rayleigh number. The four panels refer to three different salinity cases plus the density matched (heavy water) case. Each experiment is represented by a line with markers. Squares indicate an initial diameter of \qty{5.0}{\centi\meter}, triangles of \qty{8.1}{\centi\meter}, and circles of \qty{12.0}{\centi\meter}. As time progresses, the cylinder shrinks, so time direction is from higher Ra to lower Ra. The dashed lines indicate a scaling of Nu $\propto \text{Ra}^{1/3}$. The initial increase of Nusselt is explainable by the initial heat diffusion inside the ice. The decrease of the Nusselt number in the last stages of the melting in saline conditions is due to the accumulation of cold fresh water under the surface of the ice. }
    \label{fig:NuRa}
\end{figure}

The scaling of the Nusselt number against the Rayleigh number is provided in figure \ref{fig:NuRa}. The data follow a convective Nu $\propto$ Ra$^{1/3}$ scaling that is reported in literature for laminar type thermal boundary layers \citep{yamada_melting_1997,hosseini_experimental_2009,churchill_correlating_1975,wells_geophysical-scale_2008,grossmann_scaling_2000}. The data follow the 1/3 scaling relation regardless of shape or initial size. A feature appearing in all the panels is the initial increase of Nusselt number. {We believe this is due to the temperature profile in the ice becoming less steep as temperature diffuses in the solid.} \par
In the later stages of the melting process, the saline water measurements show a drastic decrease of Nusselt number. As described previously, under saline conditions, the meltwater rises and creates an insulating layer which hinders melting. In terms of the Nusselt vs Rayleigh dependence, this results in a lower Nusselt compared to the predicted 1/3 scaling. The intensity of this deviation depends on the initial size of the cylinder, and not on the Rayleigh number: at a given Rayleigh number, cylinders that have started larger will have melted a larger part of their volume, and consequently the thickness of the subsurface meltwater layer will be larger. Moreover, the mixing between the meltwater and the saline (surrounding) water depends on the salinity itself. This is visible in the early-time deviation from the scaling relation for the \qty{35}{\gram\per\liter} experiments compared to the \qty{10}{\gram\per\liter} ones. The deviation  shown in few curves in the \qty{0}{\gram\per\liter} panel, at around $10^{5}\lessapprox \text{Ra} \lessapprox 10^{6}$ can be ascribed to the transition from a convection dominated regime (Nu $\gg1$) to a diffusion dominated regime (Nu $=\mathcal{O} (1)$). The heavy water experiments (D$_2$O) show a similar Nu vs Ra scaling relation as the others. We consider the top (convex) curve less reliable because, in order to reduce the consumption of the expensive heavy water, we choose to image the melting together with the PIV data. Hence our imaging setup was not designed for contour recognition.  \par

\subsection{Plume measurements}\label{sec:03.5plume}
\begin{figure}
    \centering
    \includegraphics[width=\textwidth]{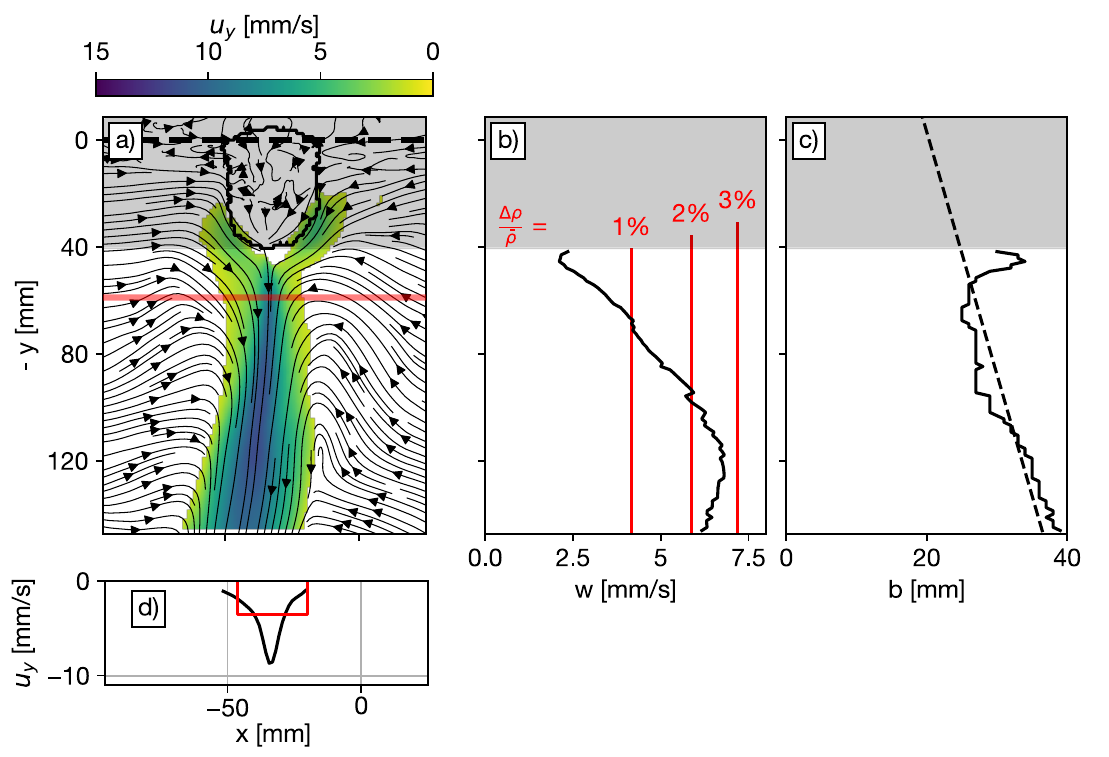}
    \caption{Characteristics of the plume generated by a melting cylinder (D$_\text{equivalent}$ = \qty{2.04}{\centi\meter}, D$_\text{initial}$ = \qty{5.0}{\centi\meter}, T$_\text{water}$ = \qty{17.9}{\celsius}$\pm$\qty{0.2}{\kelvin}, S = \qty{35}{\gram\per\liter}, Ra$_\text{initial}$ $\approx$ \num{2e8}). Panel (a) shows the thresholded downward velocity field as obtained from PIV data, with superimposed: streamlines; contour of the ice; water level (horizontal dashed line); the depth at which the profile in panel (d) has been extracted (red horizontal strip). Panel (b) shows the typical velocity of the plume as a function of depth. {The vertical red lines are the velocity that is expected when considering a constant density anomaly of 1\%, 2\%, and 3\% above the reference density.} Panel (c) shows the width of the plume with a linear fit to the data. In panels (a), (b), and (c) the dark shading refers to the region outside our region of interest (ROI). Panel (d) shows the velocity profile in the plume at the cross-section identified in panel (a). The horizontal red line is the measured plume width. }
    \label{fig:plume}
\end{figure}

From the PIV measurements, we identified the downward plume by thresholding the vertical velocity. The threshold was chosen of the same magnitude as the converging horizontal flows. Different threshold values do not produce significantly different results. We located the bottom of the ice from the contour information derived from the imaging and considered the velocity field as starting below the ice. Panel (\textit{a}) of figure \ref{fig:plume}  summarises the characteristics of the plume's velocity field for a cylinder (initial diameter \qty{50}{\milli \meter}, salinity \qty{35}{\gram\per\liter}). The behaviours described hereafter are consistently seen in all experiments (with different initial size and different salinities). \par
As seen in panel (\textit{a}) of figure \ref{fig:plume}, the actual source of the plume is spread both horizontally and vertically. We determined the velocity profile of the plume by taking the horizontal mean of the velocity field. {At a certain depth, the width of the plume was defined as the one that would conserve the downward momentum at that depth, while assuming a top-hat profile.} Panel (\textit{d}) shows a typical velocity profile, measured at the transect indicated in red in panel (\textit{a}). Panel (\textit{c}) of the figure shows that the plume width increases linearly with depth, in agreement with our line-source plume theory. By performing a linear fit between plume width and depth ($b=\alpha y$) we find an entrainment coefficient of $\alpha= 0.11$.  \par 
A precise measure of the plume density difference as a function of depth is not feasible. A reasonable estimate is obtained by assuming that at the highest point of the plume (just below the ice) the density of the plume was of saline water cooled to zero degrees Celsius. In this case, the density anomaly compared to ambient-temperature saline water is around 3\%. In panel (\textit{b}) we plot the expected plume velocities for density anomalies of 1\%, 2\%, and 3\%, calculated with equation \ref{eqn:plumeVelocity}. {We remark that the theory that we presented in Section \ref{sec:03.5plume} assumes a top-hat profile for the plume's velocity, which is to be compared with the example profile of panel (\textit{d}) of figure \ref{fig:plume}. This adds on top of the relatively short distance from the source at which our measurements are taken, which relates to the assumption of considering our plumes to be fully-developed and turbulent. With these considerations, our measurements show a marked acceleration of the plume with greater depths, and the velocity does not differ significantly from the expected value.}\par
\begin{figure}
    \centering
    \includegraphics[width=\textwidth]{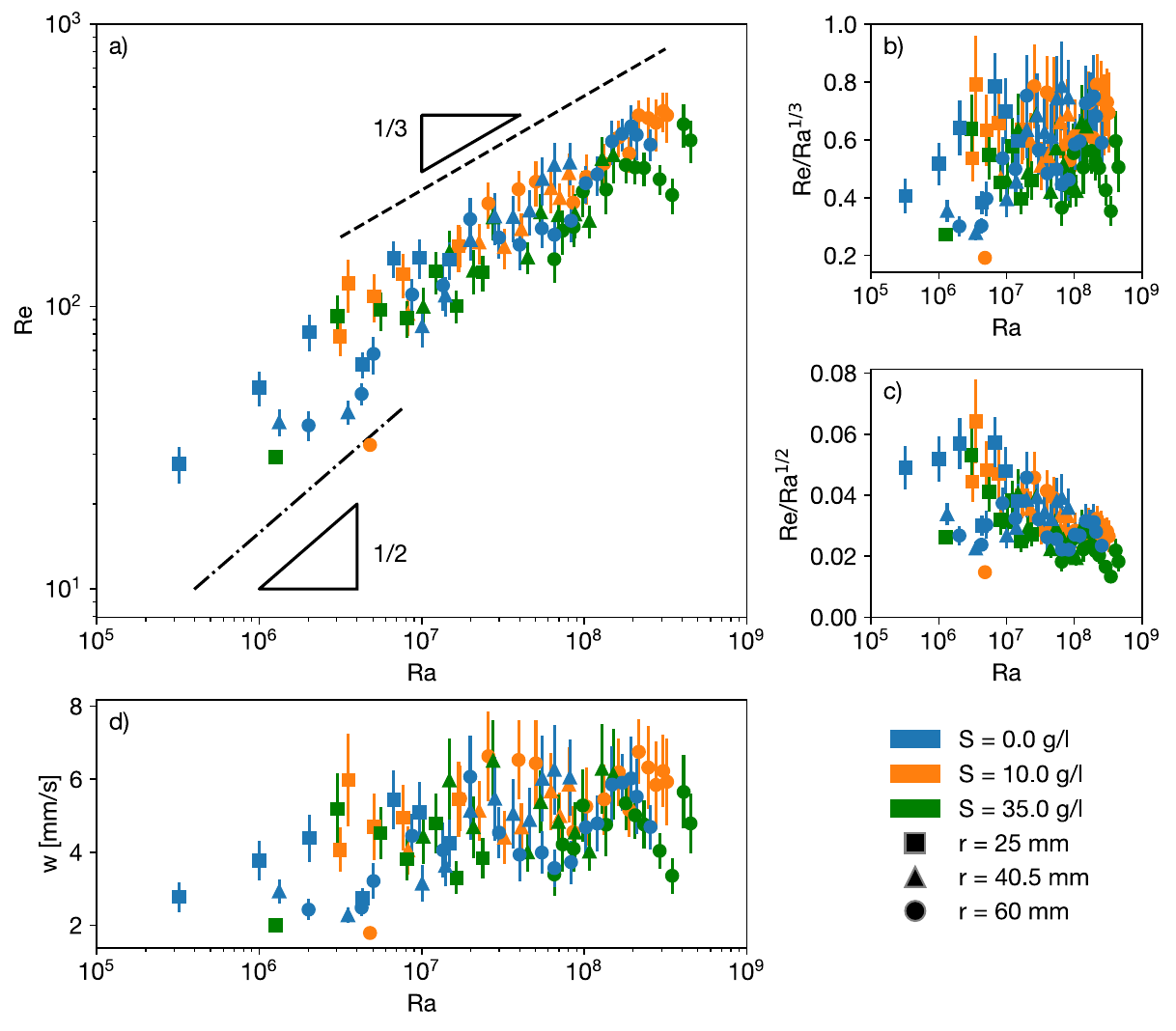}
    \caption{Reynolds number against Rayleigh number for all the experiments from our PIV data. Each point refers to a PIV dataset (acquisition time \qty{20}{\second}). The color of the marker (blue, orange or green) refers to the salinity of the surrounding water in the experiment, the shape of the marker (square, triangle, disc) to the initial radius of the cylinder. Panel (a) reports the uncompensated plot, with the dash-dotted line indicating a Re $\propto \text{Ra}^{1/2}$ scaling and the dashed line a Re $\propto \text{Ra}^{1/3}$ scaling. Panel (b) and (c) present the compensated plots for these two scaling relations. Panel (d) reports the mean downward velocity (see figure \ref{fig:plume}a) as a function of Ra. }
    \label{fig:ReRa}
\end{figure}

Panel (\textit{a}) of figure \ref{fig:ReRa} plots the Reynolds number as a function of the Rayleigh number of all of our experiments. Straight-line fits of our data are assisted with a Re $\propto$ Ra$^{1/2}$ scaling for the lower region of Rayleigh number (up to Ra $\approx 10^7$) and a Re $\propto$ Ra$^{1/3}$ scaling for higher Ra. This is supported by panels (\textit{a}), (\textit{b}) of the figure. Given that Re depends on $w$ and $R$, and Ra has its strongest dependency on $R^3$, a 1/3 scaling implies a constant downward velocity. This is confirmed in panel (\textit{d}), which shows a constant downward velocity for the range of Ra where the 1/3 scaling is valid. Indeed, a roughly constant downward velocity is predicted by equation \ref{eqn:plume}: given that the downward velocity $w$ depends only on the density difference, and the density difference varies only of a few percent throughout our experiments as a result of different salinities. The equations thus predict a velocity that varies only up to few percent, much below our experimental accuracy. The scaling relations for Re are in accordance with the GL theory \citep{grossmann_scaling_2000,grossmann_thermal_2001}, which indeed features a transition from a regime $I_u$ (dominated by boundary layer dissipation, for velocity boundary layer larger than thermal boundary layer) to a regime $IV_u$ (dominated by bulk dissipation, for velocity boundary layer larger than thermal boundary layer). \par

\begin{figure}
    \centering
    \includegraphics[width=.8\textwidth]{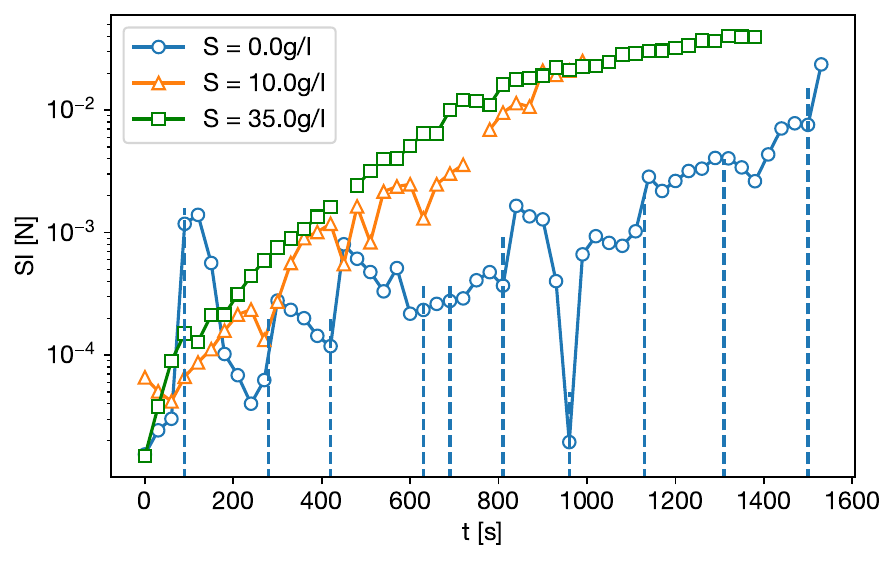}
    \caption{Temporal evolution of the stability index for three cylinders (for all, D$_\text{initial}$ = \qty{8.1}{\centi\meter}). They melted in water with salinity S = 0 (circles, T$_\text{water}$ = \qty{20.9}{\celsius}$\pm$\qty{0.2}{\kelvin}, Ra$_\text{initial}$ $\approx$ \num{5.1e7}), S = \qty{10}{\gram\per\liter} (triangles, T$_\text{water}$ = \qty{19.0}{\celsius}$\pm$\qty{0.2}{\kelvin}, Ra$_\text{initial}$ $\approx$ \num{1.7e8}) and S = \qty{35}{\gram\per\liter} (squares, T$_\text{water}$ = \qty{19.3}{\celsius}$\pm$\qty{0.2}{\kelvin}, Ra$_\text{initial}$ $\approx$ \num{6.9e8}). Vertical dashed lines indicate the times of rotation of the freshwater cylinder (the others do not rotate). For the freshwater cylinder, the stability index increases during a rotation. For the saline water cylinders, the stability index increases monotonically over time. }
    \label{fig:stability}
\end{figure}

\subsection{Time evolution of the stability of cylinders}\label{sec:03.3salt}
Figure \ref{fig:stability} reports the temporal evolution of the stability index SI, as defined in subsection \ref{sec:stabIndex}, for three cases: fresh water S=\qty{0}{\gram\per\liter}, saline water with S=\qty{10}{\gram\per\liter}, and saline water with S=\qty{35}{\gram\per\liter}. The times of rotations of the freshwater cylinder are marked as vertical dashed lines (the saline cases do not rotate). {First, we can support the statement that it is energetically favourable for the cylinder to rotate --- i.e. that a rotation lowers the gravitational potential energy of the cylinder --- by noting that in the fresh water case the stability index shows a well-defined increase across rotations}\footnote{In some rotation events, the increase in stability index is less prominent than in others. This is due to two factors. In some cases the angular rotation is very small, and the vertical displacement of the center of mass is minimal. In some other case, the cylinder ``overshoots'', surpassing the closest stable equilibrium point and reaching a further one, which may have a similar stability index.}. Second, the stability index of the saline water cases shows a clear increase over time. This is consistent with our experimental observations of the absence of rotations in the saline water case, and supports the statement that the morphology evolution progressively stabilises the cylinder. {Most notably, the evolution of the stability of the cylinder is not connected to the flow dynamics, but rather only to the balance between gravity and buoyancy.} \par

\subsection{Rotation dynamics}\label{sec:03.2rotation}
\begin{figure}
    \centering
    \includegraphics[width=\textwidth]{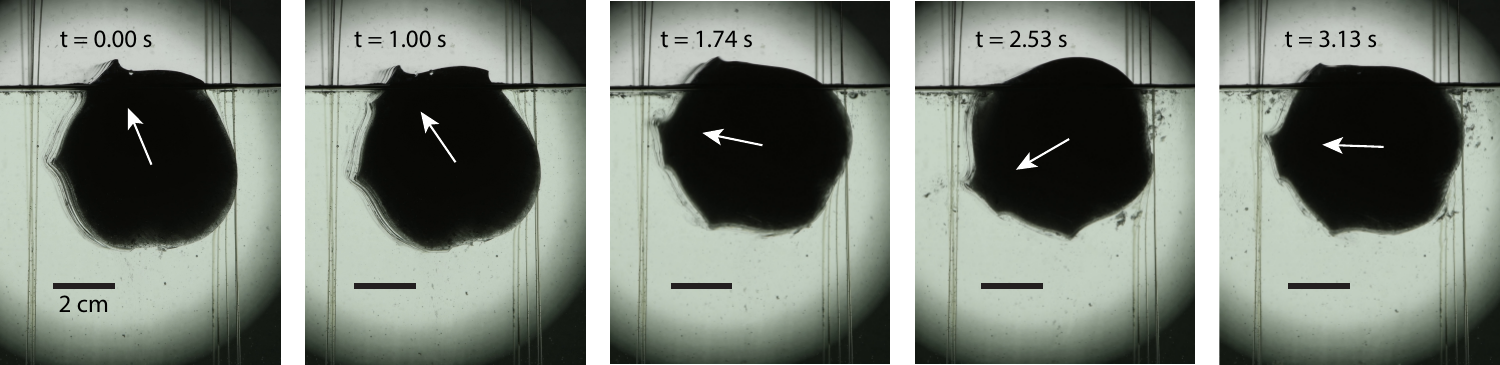}
    \caption{Images of an ice cylinder (d$_\text{equivalent}$ = \qty{5.8}{\centi\meter}, d$_\text{initial}$ = \qty{8.1}{\centi\meter}, T$_\text{fluid}$ = \qty{19.6}{\celsius}$\pm$\qty{0.2}{\kelvin}, S = \qty{0.0}{\gram\per\liter}, Ra $\approx$ \num{1.6e7}) performing a counterclockwise rotation, shown as an example of the mechanism. The white arrow is added to the image to help the reader track the motion of the cylinder. The images are taken on a later stage of the melting, and the ice is already not circular anymore. The vertical lines visible in the image are the nylon wires to keep the ice in place. The ice is rotating and then oscillating around the new stable equilibrium.  }\label{fig:shots}
\end{figure}

\begin{figure}
    \centering
    \includegraphics[width=.7\textwidth]{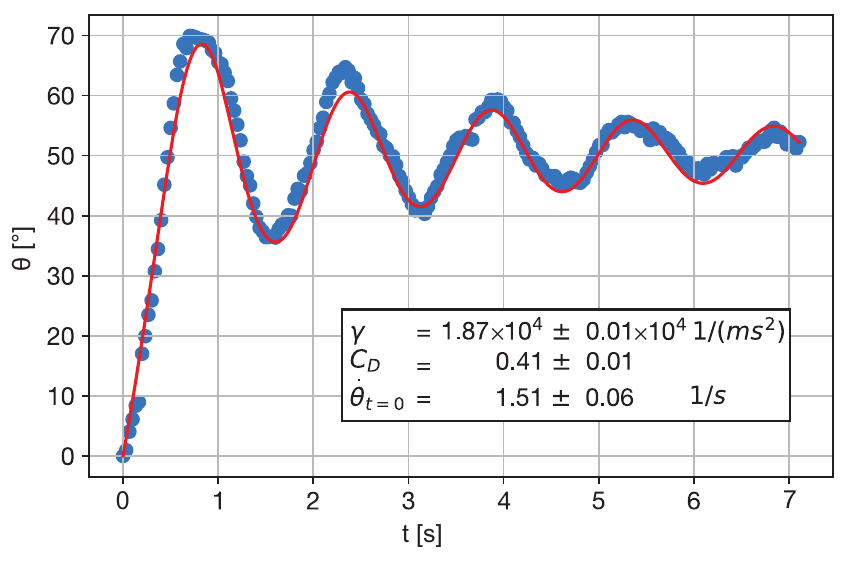}
    \caption{Best fit of the solution of equation \ref{eqn:fittedRotation} to the data of a rotational oscillation of a cylinder (D$_\text{equivalent}$ = \qty{5.9}{\centi\meter}, D$_\text{initial}$ = \qty{8.1}{\centi\meter}, T$_\infty$ = \qty{19.3}{\celsius}$\pm$\qty{0.2}{\kelvin}, S = \qty{0.0}{\gram\per\liter}, Ra$_\text{initial}$ $\approx$ \num{4e7}). The blue dots refer to the angle of rotation around the center of mass, where the zero is set to the initial position of the ice. The red line is a the result of fitting the parameters of equation \ref{eqn:fittedRotation} ($\gamma$, $C_D$, and $\dot{\theta}_{t=0}$) to match the experimental data. }
    \label{fig:rotation}
\end{figure}

A sequence of images of a typical rotation of a cylinder melting in freshwater is shown in figure \ref{fig:shots}. Figure \ref{fig:rotation} shows the angle of orientation with respect to the vertical of an ice cylinder during a capsizing event. We fit the data with equation \ref{eqn:fittedRotation}, with $\gamma$, $C_D$, and $\dot{\theta}(t=0)$ being the fitting parameters. Any experimental error in the measurement of $d(\theta)$ or inaccuracy in the density of the ice or water are accounted for in the errors of $\gamma$ and $C_D$. A first estimate for $C_D$ is in the range 0.5--1, though a proper estimation is impossible due to the unique shape that the ice has. {The equivalent radius $r$ of the cylinder can be evaluated from the contour ((\num{29.5} $\pm$ \num{3}) \unit{\milli\meter} in the case of figure \ref{fig:rotation}), leading to a $\gamma$ equal to \mbox{(\num{2.1} $\pm$ \num{0.02}) $\times$ 10$^4$ \unit{1\per (\meter \second \squared)}}. The fit returns a $C_D$ of \qty{0.41} and a $\gamma$ of \mbox{(\num{1.9} $\pm$ \num{0.01}) $\times$ 10$^4$ \unit{1\per (\meter \second \squared)}}. We believe that the mismatch in the $\gamma$ value is mainly due to the assumption of a constant cross section of the cylinder along its axis, which is directly influencing the accuracy of the radius and $d(\theta)$.} \par

\section{Extension to general two-layer stratifications}\label{sec:twolayer}
In our saline water experiments, the melting cylinder created a two-layer density stratification in the water, which was initially at constant density. In nature, however, there are situations where floating ice melts in already-stratified water, like in ice melanges (see \citet{burton_quantifying_2018} and references therein), or pack ice \citep{rothrock_mechanical_1975}, or in fjords \citep{fitzmaurice_effect_2016,jackson_externally_2014}. A common feature of these examples is that in an enclosed basin multiple icebergs melt at the same time. The case of multiple ice blocks melting at the same time goes beyond the scope of this work, but we can easily access the case of an initial two-layer density stratification in the water. \par
We have conducted two sets of experiments. All of them with the \qty{81}{\milli \meter} diameter cylinders at an initial temperature of \qty{-16}{\celsius}. The first set was such that the fresh water layer left by a molten cylinder was used as the initial condition for the subsequent one. In our tank, the volume of a cylinder corresponds to a water height of approximately \qty{5}{\milli \meter}. Thus, the first cylinder melted in saline water (S $=\qty{35}{\gram \per\liter}$), the second one with a fresh water layer on top of \qty{5}{\milli \meter}, the third one with \qty{10}{\milli \meter}, the fourth with \qty{15}{\milli \meter}, and so on. The temperature of the top layer was not controlled, but it was determined by the influx of meltwater at \qty{0}{\celsius}, and the heat transfers with the air above and the saline water below. The second set of experiments was done with the same layer thicknesses, but the fresh water layer was created every time with water at \qty{24}{\celsius}. The sharp stratification was created by carefully pouring water on a floating sponge, that would break the vertical momentum of the poured water. In this configuration, the meltwater accumulates in between the bottom saline layer and the top warm freshwater layer (see panel (c) of figure \ref{fig:boundaryLayers}). \par
\begin{figure}
\centering
\sidesubfloat[]{\label{fig:firstSetLayered}\includegraphics[width=0.9\textwidth]{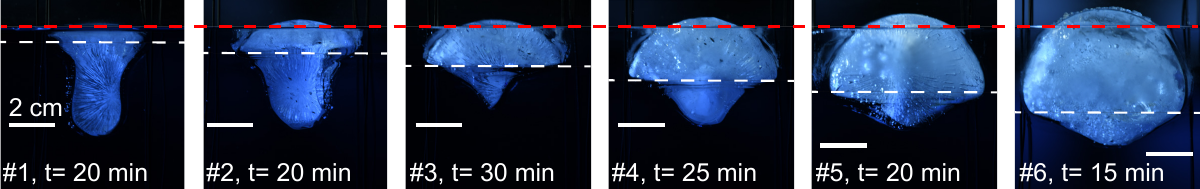}}
\vspace{5mm}
\sidesubfloat[]{\label{fig:secondSetLayered}\includegraphics[width=0.7\linewidth]{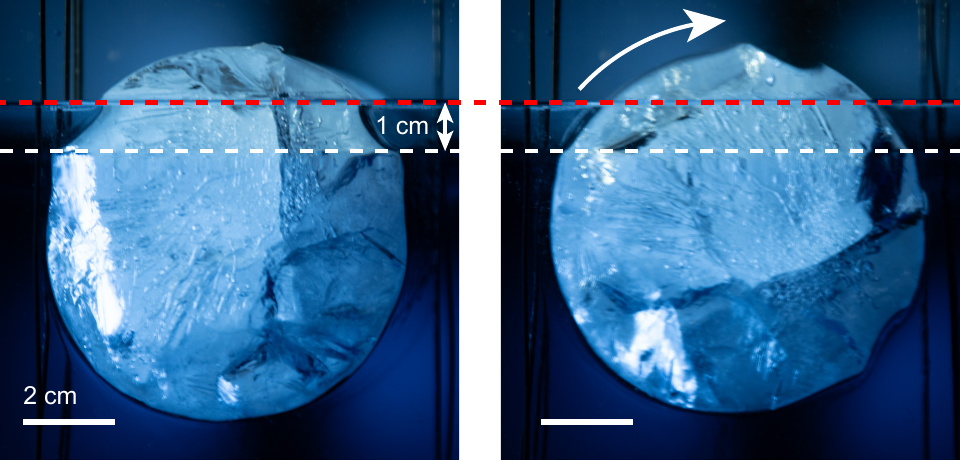}}
\caption{(a) Snapshots of subsequently melting ice cylinders (for all, D$_\text{initial}$ = \qty{8.1}{\centi\meter}). The red dashed line indicates the water surface. {The first cylinder was melted in saline water (S = \qty{35}{\gram \per \liter}, T = \qty{19.8}{\celsius}$\pm$\qty{0.2}{\kelvin}).} The time indicated at the bottom of each panel is the time elapsed from the beginning of each experiments. The number following the hash symbol is the progressive number of experiment. The meltwater accumulating on top of the denser saline water created a two-layer stratification, with the top layer being fresh and cold and the bottom layer being the original ambient water. The white dashed line indicates the extent of the meltwater layer. {Each following experiment was melted in the stratification as left by the previous experiment. }The blue shade on the picture is due to food colouring used to chromatically distinguish the bottom layer from the top one. \\
(b) Two shots of an ice cylinder (D$_\text{initial}$ = \qty{8.1}{\centi\meter}) melting in two-layered water, with the top layer being fresh and the bottom layer being at salinity S = \qty{35}{\gram \per \liter}. The temperatures of the two layers were T$_\text{top}$ = \qty{20.6}{\celsius}, T$_\text{bottom}$ = \qty{24.2}{\celsius}. The two shots correspond to two instants before and after a rotation of the cylinder. The blue shade on the picture is due to food colouring used to chromatically distinguish the bottom layer from the top one. The red dashed line is the water surface, the white dashed line is the pycnocline. The thickness of the top layer is the lowest for which we can ensure a visually sharp distinction between the two layers.}
\end{figure}

As visible in figure \ref{fig:firstSetLayered}, all the first set of experiments resulted in rotationally stable cylinders. In fact, the melting was primarily happening in the bottom saline layer, and the resulting shape was stable. In addition to this, the melting time increased substantially, given that --- after the melting of the lower part of the cylinder --- the remaining part was immersed in cold freshwater. On the contrary, all the second set of experiments resulted in unstable cylinders (an example visible in figure \ref{fig:secondSetLayered}). Note here that it was experimentally impossible to artificially create a sharp layer of warm freshwater thinner than \qty{5}{\milli \meter}. Although the morphological changes of these experiments differ from the ones seen in the full fresh water, the reasons for the rotational instability still lie in the horizontal displacement of buoyancy and gravity. \par

\floatsetup[figure]{style=plain,subcapbesideposition=top}

\section{Conclusions and outlook}\label{sec:04conclusions}
We have conducted melting experiments in a \qty{115}{\liter} aquarium filled with quiescent water, where ice cylinders were put to float with their axes perpendicular to gravity. We have changed both the density of the ambient water and that of the melting object. The salinity of the water was changed from fresh to \qty{10}{\gram\per\liter} to oceanic salinity (S = \qty{35}{\gram\per\liter}). We performed a set of experiments with deuterium oxide (heavy water) to explore differently buoyant cylinders. We monitored the morphological evolution and vertical orientation of the cylinders and imaged the convective flow around the cylinder. {Our work expands the fundamental research on floating, melting objects. This is of particular significance, given that, in several natural and industrial circumstances, floating ice particles are not fixed in place.} \par
The interaction of the water with the cold cylinder gives rise to density anomalies which drive convective flows in the tank. According to its own density anomaly, the meltwater can either mix or shear with the outer flow. {We identified the key physical mechanisms that determine the morphological evolution of the cylinders. Salinity adds a third (solutal) boundary layer to the existing two (thermal and momentum), with the three sharing no common length scale. For sufficiently high ambient water salinities, the flow in the vicinity of the ice can transition from being bidirectional closer to the bottom of the cylinder, to unidirectional closer to the air--water interface. The local melting of the ice by the flow is, consequently, drastically different in saline waters as compared to fresh waters. }\par
We found that the Nusselt number (non-dimensional heat transfer) against the Rayleigh number scales as Nu $\propto$ Ra$^{1/3}$. This scaling is independent of size and salinity, and aligns with previous literature findings. \par
We studied the sinking plume that develops below the cylinders and compared its dynamics with that of line-source plume. Despite being far from an ideal plume, the velocity and width profiles are consistent with the predictions with the theory for a line source plume. The Reynolds number of the velocity of the line source plume was plotted against the cylinder's Rayleigh number. {The data collapses on a line independently of size and salinity, and aligns with known theory for plumes (linearly increasing plume width with depth and constant plume velocity) and convective heat transfers (Re $\propto$ Ra$^{1/2}$ for Ra $< \mathcal{O}(10^7)$ and Re $\propto$ Ra$^{1/3}$ for Ra $> \mathcal{O}(10^7)$).} \par
Cylinders melting in fresh water experience repeated capsizing events, which result from an asymmetric meltrate between the part exposed to air and the part exposed to water. On the contrary, cylinders melting in saline water have a cross-section that is highly stable against rotations, hence do not capsize. Based on Newton's second law for rotating objects, we developed a measure of the stability of the cylinders, which qualitatively describes the observed phenomena. We modelled the oscillations that occur after a capsize via a damped non-linear oscillator model, which fits our data satisfactorily. \par
Finally, we conducted experiments in two-layer stratified water, and showed that the ice is stable when the melt rate is higher in the bottom layer, and unstable when the meltrate is higher in the upper layer. {We argue that our experimental findings in the laboratory aquarium are qualitatively connected to several natural circumstances where icebergs melt in stratified waters, with the substantial difference that real icebergs overturn in saline waters, while our ice cylinders do not.} {In fact, a few caveats apply to this analogy. }\par
Firstly, the density anomaly that we observed in the laboratory is of the same order as the density difference between fresh and salty water (\qtyrange{25}{30}{\gram\per\liter}), while in observations they report much lower anomalies, of the order of \qty{0.5}{\gram\per\liter}. This is due to the different melting timescales between our experiments and real icebergs. In fact, the ocean temperature at the poles can be as low as zero degrees, and the melting process in such conditions is greatly slowed down, hence the mixing time scales are close to the melting time scale. On the contrary, our ice melts much quicker than the mixing timescales.\par
Secondly, the typical thickness of the naturally-occurring layer is both hard to determine and impossible to repeat in a laboratory. \citet{yankovsky_surface_2014} report significant density stratifications up to \qty{-40}{\meter} of depth, but they report lack of data in the upper \qty{14}{\meter} layer. Direct measurements of icebergs' keel depth are not usually carried out during oceanographic cruises, but \citet{dowdeswell_keel_2007} proposed an estimate of the keel's depth based on satellite-altimetry data. Their frequency distribution presents three peaks: one at \qtyrange{150}{200}{\meter}, one around \qtyrange{250}{300}{\meter}, and one at \qtyrange{550}{600}{\meter}. Detailed values depend on the region of survey. Hence, with the information of a typical plume depth in the order of tens of meters, and a typical iceberg keel depth of the order of hundreds of meter, we conclude that our centimeter thick layer with a \qty{10}{\centi\meter}-thick ice is proportionally not too far off from typical natural occurrence.\par
{Thirdly, a number of mechanisms occur only in nature and not in our laboratory, such as wave erosion and solar radiation. These two typically act on the top layer of an iceberg, resulting in an increased ablation rate of the ice close to the surface of the water. This ablation rate is significantly different from the one of our ice cylinders that melt in well-mixed saline water, but is similar to the one of cylinders melting in stratified water (with the top part melting quicker than the bottom part). For these reasons, the capsizing behaviour of these ice cylinders can be connected with the one of icebergs. Another feature that only appears in the natural icebergs is the presence of (unevenly distributed) sediments and large pockets of air and air bubbles. On top of altering the ablation dynamics, these also generate density inhomogeneities that could determine a significant difference in the stability of real icebergs compared to our cylinders. }\par
In our future experiments, we will vary the temperature of the surrounding water, to lower the corresponding contribution to the melt rate. 

\backsection[Acknowledgements]{The authors wish to acknowledge: Gert-Wim Bruggert, Martin Bos, and Thomas Zijlstra for technical support; Hannah Brugge for her preliminary measurements. We wish to thank Dr. Claudia Cenedese for fruitful discussions.}

\backsection[Funding]{This work was financially supported by the European Union (ERC, MeltDyn, 101040254 and ERC, MultiMelt, 101094492).}

\backsection[Declaration of interests]{The authors report no conflict of interest.}

\backsection[Data availability statement]{The contours of the cylinders are openly available in 4TU.ResearchData at \href{http://doi.org/10.4121/bcb292f1-7a51-4f8d-a175-4a04a3d02e4c}{http://doi.org/10.4121/bcb292f1-7a51-4f8d-a175-4a04a3d02e4c}. The codes for data analysis are available at \href{https://github.com/ebellincioni/floatingIce}{https://github.com/ebellincioni/floatingIce}.}

\backsection[Authors ORCIDs]{Edoardo Bellincioni, \href{https://orcid.org/0009-0002-3016-6135}{0009-0002-3016-6135}; Detlef Lohse, \href{https://orcid.org/0000-0003-4138-2255}{0000-0003-4138-2255}, Sander G. Huisman, \href{https://orcid.org/0000-0003-3790-9886}{0000-0003-3790-9886}}

\appendix 
\section{Experiments with deuterium oxide}\label{app:a}
\begin{figure}
    \centering
    \includegraphics[width=1\textwidth]{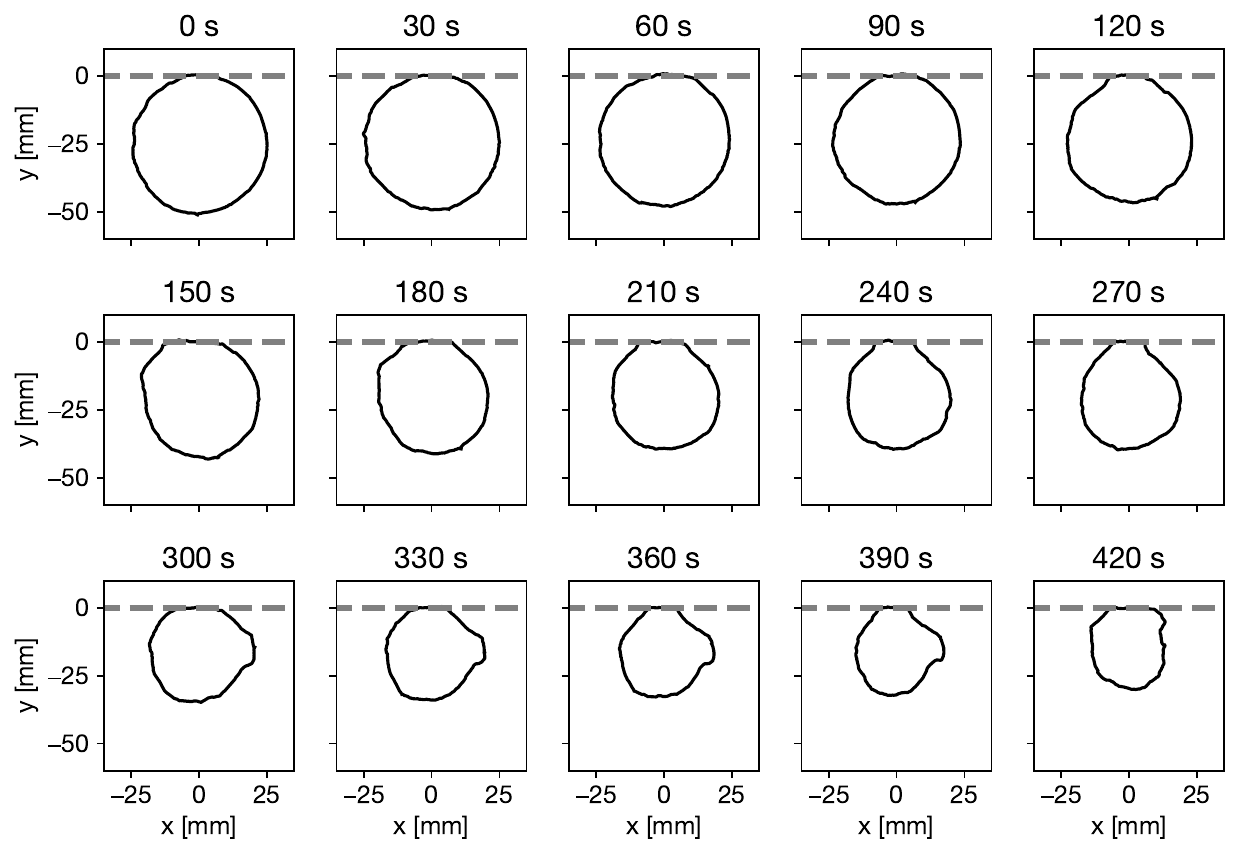}
    \caption{Shape evolution of a density-matched ice cylinder (mixture of D$_\text{2}$O and H$_\text{2}$O, D$_\text{initial}$ = \qty{5.0}{\centi\meter}, Ra$_\text{initial}$ $\approx$ \num{4.9e8}) melting in fresh water (T$_\infty$ = \qty{18.7}{\celsius}$\pm$\qty{0.2}{\kelvin}, S = \qty{0.0}{\gram\per\liter}). Each panel is labelled with the time from the beginning of the experiment. The grey dashed line represents the air-water interface.}
    \label{fig:heavyWater}
\end{figure}

To investigate ice which does not expose parts to air, but still floats, we prepared a mixture of D$_\text{2}$O and H$_\text{2}$O such that its ice would have the same density of water. The inevitable presence of bubbles (estimated to be 1\% in volume) makes the ice to float slightly below the water surface, without breaking surface tension, see figure \ref{fig:heavyWater}. The morphology of this ice cannot evolve due to the different heat transfers between parts exposed to water and parts exposed to air, but only due to the flow sculpting the ice. The universality of the mechanisms proposed for non-rotating ice is supported by the similarities between the first panels of figure \ref{fig:heavyWater} and the non-rotating ice in figure \ref{fig:additionalShapes}. In later stages (e.g. between \qty{120}{\second} and \qty{150}{\second}), the ice capsizes. Compared to the previously described case, however, here the instability is caused by the reduced buoyancy of the subsurface ice, and not by the upward vertical displacement of the center of mass of the ice. \par

\section{Explanation of stability index}\label{app:b}
\begin{figure}
    \centering
    \includegraphics[width=0.7\linewidth]{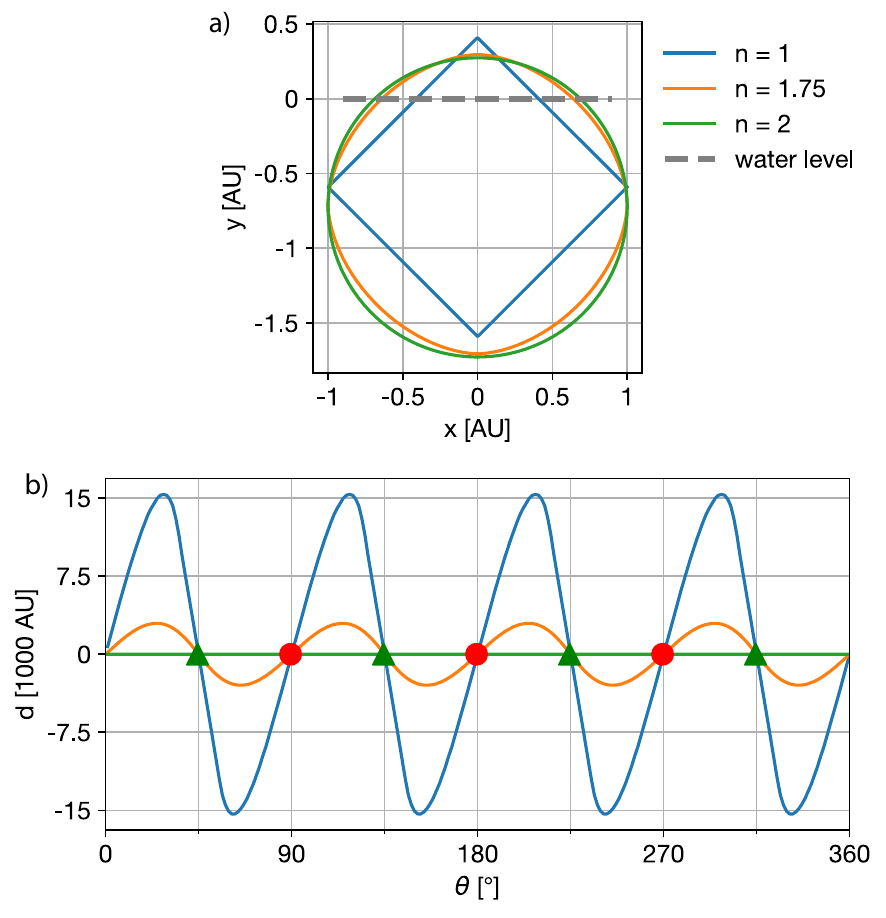}
    \caption{{Examples of shapes with different stability characteristics. Panel (a) shows the shapes, defined through the mathematical expression for a superellipse. The shapes are assumed to ``float'' in water, with the same density of ice. The horizontal dashed line indicates the water level. The dimensions of the shape are in arbitrary units AU.  Panel (b) reports the variation of the horizontal distance between centre of mass and centre of buoyancy throughout a \qty{360}{\degree} rotation of the shapes. The shapes of panel (a) are at $\theta =0$. Stable equilibrium points are indicated with green triangles, while unstable with red circles. }}
    \label{fig:appStabIndex}
\end{figure}

{To help the reader understand the two mechanisms that determine the stability of floating cylinders, let us consider the stability of three floating superellipses. The superellipse is given by the following relation}
\begin{equation}
    {
    \bigg |\frac{x}{a}\bigg|^n+\bigg |\frac{y}{b}\bigg|^n\leq1 \,\,. 
    }
\end{equation}
{Fixing $a=b=1$, for $n=1$ the shape is a square, for $n=2$, the shape is a disk, for $n\in[1,2]$ is a square with round sides, see top panel of figure \ref{fig:appStabIndex}. The bottom panel of the same figure reports the variation of horizontal distance $d$ between the centre of mass (com) and centre of buoyancy (cob) throughout a 360$^\circ$ rotation. In the case of the disk, the com and cob are always vertically aligned, and $d$ is identical to zero. For the other two shapes, $d$ oscillates between positive and negative values, and its zeros correspond to equilibrium points. The angular distance between equilibrium points is the same in the two cases (equal to \qty{45}{\degree}), but the amplitude of the oscillations is not. The magnitude of the force required to have the shape capsize depends on $d$, and it's consequently lower in the case of $n=1.75$ compared to $n=1$, despite the difference in area. At the same time, any perturbation to a stable equilibrium point that will not tilt the shape enough to reach the closest unstable equilibrium point (less than \qty{45}{\degree}, in the examples) will not have the ice capsize, but oscillate back to the original stable equilibrium point. }\par
{For a different shape (e.g. with a different number of vertices), both the angular distance between equilibrium points and the amplitude of the $d$ oscillations will change. }

\vspace{20mm}


\begin{thebibliography}{48}
\expandafter\ifx\csname natexlab\endcsname\relax\def\natexlab#1{#1}\fi
\def\au#1{#1} \def\ed#1{#1} \def\yr#1{#1}\def\at#1{#1}\def\jt#1{\textit{#1}} \def\bt#1{#1}\def\bvol#1{\textbf{#1}} \def\vol#1{#1} \def\pg#1{#1} \def\publ#1{#1}\def\arxiv#1{#1}\def\org#1{#1}\def\st#1{\textit{#1}}

\bibitem[Bigg {\em et~al.\/}(2014)Bigg, Wei, Wilton, Zhao, Billings, Hanna \& Kadirkamanathan]{bigg_century_2014}
{\sc \au{Bigg, G.~R.}, \au{Wei, H.~L.}, \au{Wilton, D.~J.}, \au{Zhao, Y.}, \au{Billings, S.~A.}, \au{Hanna, E.} \& \au{Kadirkamanathan, V.}} \yr{2014}  \at{A century of variation in the dependence of {Greenland} iceberg calving on ice sheet surface mass balance and regional climate change}.  \jt{Proc R. Soc. A.}  \bvol{470}~(2166),  \pg{20130662}.

\bibitem[Bonnet {\em et~al.\/}(2020)Bonnet, Yastrebov, Queutey, Leroyer, Mangeney, Castelnau, Sergeant, Stutzmann \& Montagner]{bonnet_modelling_2020}
{\sc \au{Bonnet, P}, \au{Yastrebov, V~A}, \au{Queutey, P}, \au{Leroyer, A}, \au{Mangeney, A}, \au{Castelnau, O}, \au{Sergeant, A}, \au{Stutzmann, E} \& \au{Montagner, J-P}} \yr{2020}  \at{Modelling capsizing icebergs in the open ocean}.  \jt{Geophys. J. Int.}  \bvol{223}~(2),  \pg{1265--1287}.

\bibitem[Burton {\em et~al.\/}(2012)Burton, Amundson, Abbot, Boghosian, Cathles, Correa-Legisos, Darnell, Guttenberg, Holland \& MacAyeal]{burton_laboratory_2012}
{\sc \au{Burton, J.~C.}, \au{Amundson, J.~M.}, \au{Abbot, D.~S.}, \au{Boghosian, A.}, \au{Cathles, L.~M.}, \au{Correa-Legisos, S.}, \au{Darnell, K.~N.}, \au{Guttenberg, N.}, \au{Holland, D.~M.} \& \au{MacAyeal, D.~R.}} \yr{2012}  \at{Laboratory investigations of iceberg capsize dynamics, energy dissipation and tsunamigenesis}.  \jt{J. Geophys. Res.}  \bvol{117}~(F1).

\bibitem[Burton {\em et~al.\/}(2018)Burton, Amundson, Cassotto, Kuo \& Dennin]{burton_quantifying_2018}
{\sc \au{Burton, J.~C.}, \au{Amundson, J.~M.}, \au{Cassotto, R.}, \au{Kuo, C.} \& \au{Dennin, M.}} \yr{2018}  \at{Quantifying flow and stress in ice m{\'e}lange, the world{\textquoteright}s largest granular material}.  \jt{Proc. Natl. Acad. Sci. U.S.A.}  \bvol{115}~(20),  \pg{5105--5110}.

\bibitem[Carey \& Gebhart(1982)]{carey_transport_1982}
{\sc \au{Carey, Van~P.} \& \au{Gebhart, Benjamin}} \yr{1982}  \at{Transport near a vertical ice surface melting in saline water: experiments at low salinities}.  \jt{J. Fluid Mech.}  \bvol{117},  \pg{403--423}.

\bibitem[Cenedese \& Linden(2014)]{cenedese_entrainment_2014}
{\sc \au{Cenedese, C.} \& \au{Linden, P.~F.}} \yr{2014}  \at{Entrainment in two coalescing axisymmetric turbulent plumes}.  \jt{J. Fluid Mech.}  \bvol{752},  \pg{R2}.

\bibitem[Cenedese \& Straneo(2023)]{cenedese_icebergs_2023}
{\sc \au{Cenedese, C.} \& \au{Straneo, F.}} \yr{2023}  \at{Icebergs {Melting}}.  \jt{Annu. Rev. Fluid Mech.}  \bvol{55}~(1),  \pg{377--402}.

\bibitem[Churchill \& Chu(1975)]{churchill_correlating_1975}
{\sc \au{Churchill, S.~W.} \& \au{Chu, H. H.~S.}} \yr{1975}  \at{Correlating equations for laminar and turbulent free convection from a horizontal cylinder}.  \jt{Int. J. Heat Mass Transf.}  \bvol{18}~(9),  \pg{1049--1053}.

\bibitem[Cohen {\em et~al.\/}(2020)Cohen, Berhanu, Derr \& Courrech Du~Pont]{cohen_buoyancy-driven_2020}
{\sc \au{Cohen, C.}, \au{Berhanu, M.}, \au{Derr, J.} \& \au{Courrech Du~Pont, S.}} \yr{2020}  \at{Buoyancy-driven dissolution of inclined blocks: {Erosion} rate and pattern formation}.  \jt{Phys. Rev. Fluids}  \bvol{5}~(5),  \pg{053802}.

\bibitem[Davies~Wykes {\em et~al.\/}(2018)Davies~Wykes, Huang, Hajjar \& Ristroph]{davies_wykes_self-sculpting_2018}
{\sc \au{Davies~Wykes, M.~S.}, \au{Huang, J.~Mac}, \au{Hajjar, G.~A.} \& \au{Ristroph, L.}} \yr{2018}  \at{Self-sculpting of a dissolvable body due to gravitational convection}.  \jt{Phys. Rev. Fluids}  \bvol{3}~(4),  \pg{043801}.

\bibitem[Dorbolo {\em et~al.\/}(2016)Dorbolo, Adami, Dubois, Caps, Vandewalle \& Darbois-Texier]{dorbolo_rotation_2016}
{\sc \au{Dorbolo, S.}, \au{Adami, N.}, \au{Dubois, C.}, \au{Caps, H.}, \au{Vandewalle, N.} \& \au{Darbois-Texier, B.}} \yr{2016}  \at{Rotation of melting ice disks due to melt fluid flow}.  \jt{Phys. Rev. E}  \bvol{93}~(3),  \pg{033112}.

\bibitem[Dowdeswell \& Bamber(2007)]{dowdeswell_keel_2007}
{\sc \au{Dowdeswell, J.A.} \& \au{Bamber, J.L.}} \yr{2007}  \at{Keel depths of modern {Antarctic} icebergs and implications for sea-floor scouring in the geological record}.  \jt{Mar. Geol.}  \bvol{243}~(1-4),  \pg{120--131}.

\bibitem[FitzMaurice {\em et~al.\/}(2017)FitzMaurice, Cenedese \& Straneo]{fitzmaurice_nonlinear_2017}
{\sc \au{FitzMaurice, A.}, \au{Cenedese, C.} \& \au{Straneo, F.}} \yr{2017}  \at{Nonlinear response of iceberg side melting to ocean currents}.  \jt{Geophysical Research Letters}  \bvol{44}~(11),  \pg{5637--5644}, \_eprint: https://onlinelibrary.wiley.com/doi/pdf/10.1002/2017GL073585.

\bibitem[FitzMaurice {\em et~al.\/}(2016)FitzMaurice, Straneo, Cenedese \& Andres]{fitzmaurice_effect_2016}
{\sc \au{FitzMaurice, A.}, \au{Straneo, F.}, \au{Cenedese, C.} \& \au{Andres, M.}} \yr{2016}  \at{Effect of a sheared flow on iceberg motion and melting}.  \jt{Geophys. Res. Lett.}  \bvol{43}~(24),  \pg{12,520--12,527}.

\bibitem[Fukusako {\em et~al.\/}(1992)Fukusako, Tago, Yamada, Kitayama \& Watanabe]{fukusako_melting_1992}
{\sc \au{Fukusako, S.}, \au{Tago, M.}, \au{Yamada, M.}, \au{Kitayama, K.} \& \au{Watanabe, C.}} \yr{1992}  \at{Melting {Heat} {Transfer} {From} a {Horizontal} {Ice} {Cylinder} {Immersed} in {Quiescent} {Saline} {Water}}.  \jt{J. Heat Transf.}  \bvol{114}~(1),  \pg{34--40}.

\bibitem[Grafsr{\o}nningen \& Jensen(2012)]{grafsronningen_simultaneous_2012}
{\sc \au{Grafsr{\o}nningen, S.} \& \au{Jensen, A.}} \yr{2012}  \at{Simultaneous {PIV}/{LIF} measurements of a transitional buoyant plume above a horizontal cylinder}.  \jt{Int. J. Heat Mass Transf.}  \bvol{55}~(15-16),  \pg{4195--4206}.

\bibitem[Grafsr{\o}nningen \& Jensen(2017)]{grafsronningen_large_2017}
{\sc \au{Grafsr{\o}nningen, S.} \& \au{Jensen, A.}} \yr{2017}  \at{Large eddy simulations of a buoyant plume above a heated horizontal cylinder at intermediate {Rayleigh} numbers}.  \jt{Int. J. Heat Mass Transf.}  \bvol{112},  \pg{104--117}.

\bibitem[Grafsr{\o}nningen {\em et~al.\/}(2011)Grafsr{\o}nningen, Jensen \& Anders Pettersson~Reif]{grafsronningen_piv_2011}
{\sc \au{Grafsr{\o}nningen, S.}, \au{Jensen, A.} \& \au{Anders Pettersson~Reif, B.}} \yr{2011}  \at{{PIV} investigation of buoyant plume from natural convection heat transfer above a horizontal heated cylinder}.  \jt{Int. J. Heat Mass Transf.}  \bvol{54}~(23-24),  \pg{4975--4987}.

\bibitem[Grossmann \& Lohse(2000)]{grossmann_scaling_2000}
{\sc \au{Grossmann, S.} \& \au{Lohse, D.}} \yr{2000}  \at{Scaling in thermal convection: a unifying theory}.  \jt{J. Fluid Mech.}  \bvol{407},  \pg{27--56}.

\bibitem[Grossmann \& Lohse(2001)]{grossmann_thermal_2001}
{\sc \au{Grossmann, Siegfried} \& \au{Lohse, Detlef}} \yr{2001}  \at{Thermal {Convection} for {Large} {Prandtl} {Numbers}}.  \jt{Phys. Rev. Lett.}  \bvol{86}~(15),  \pg{3316--3319}, publisher: American Physical Society.

\bibitem[Helly {\em et~al.\/}(2011)Helly, Kaufmann, Stephenson \& Vernet]{helly_cooling_2011}
{\sc \au{Helly, J.~J.}, \au{Kaufmann, R.~S.}, \au{Stephenson, G.~R.} \& \au{Vernet, M.}} \yr{2011}  \at{Cooling, dilution and mixing of ocean water by free-drifting icebergs in the {Weddell} {Sea}}.  \jt{Deep-sea Res PT II}  \bvol{58}~(11-12),  \pg{1346--1363}.

\bibitem[Hester {\em et~al.\/}(2021)Hester, McConnochie, Cenedese, Couston \& Vasil]{hester_aspect_2021}
{\sc \au{Hester, Eric~W.}, \au{McConnochie, Craig~D.}, \au{Cenedese, Claudia}, \au{Couston, Louis-Alexandre} \& \au{Vasil, Geoffrey}} \yr{2021}  \at{Aspect ratio affects iceberg melting}.  \jt{Phys. Rev. Fluids}  \bvol{6}~(2),  \pg{023802}.

\bibitem[Hosseini \& Rahaeifard(2009)]{hosseini_experimental_2009}
{\sc \au{Hosseini, R.} \& \au{Rahaeifard, M.}} \yr{2009}  \at{Experimental {Investigation} and {Theoretical} {Modeling} of {Ice}-{Melting} {Processes}}.  \jt{Exp. Heat Trans.}  \bvol{22}~(3),  \pg{144--162}.

\bibitem[Hult \& Ostrander(1973)]{hult_antarctic_1973}
{\sc \au{Hult, John~L.} \& \au{Ostrander, Neill~C.}} \yr{1973}  \bt{Antarctic {Icebergs} as a {Global} {Fresh} {Water} {Resource}}. {\em Tech. Rep.\/}.  \org{RAND Corporation}.

\bibitem[Jackson {\em et~al.\/}(2014)Jackson, Straneo \& Sutherland]{jackson_externally_2014}
{\sc \au{Jackson, R.~H.}, \au{Straneo, F.} \& \au{Sutherland, D.~A.}} \yr{2014}  \at{Externally forced fluctuations in ocean temperature at {Greenland} glaciers in non-summer months}.  \jt{Nat. Geosci.}  \bvol{7}~(7),  \pg{503--508}.

\bibitem[Johnson {\em et~al.\/}(2023)Johnson, Zhang, Kim, Weady \& Ristroph]{johnson_poster_2023}
{\sc \au{Johnson, Bobae}, \au{Zhang, Steven}, \au{Kim, Alison}, \au{Weady, Scott} \& \au{Ristroph, Leif}} \yr{2023} Poster: {Lab} icebergs melt down and flip out.  \bt{In {\em 76th {Annual} {Meeting} of the {APS} {Division} of {Fluid} {Dynamics} - {Gallery} of {Fluid} {Motion}\/}}.  \publ{Washington, DC: American Physical Society}.

\bibitem[Josberger \& Martin(1981)]{josberger_laboratory_1981}
{\sc \au{Josberger, E.~G.} \& \au{Martin, S.}} \yr{1981}  \at{A laboratory and theoretical study of the boundary layer adjacent to a vertical melting ice wall in salt water}.  \jt{J. Fluid Mech.}  \bvol{111}~(-1),  \pg{439}.

\bibitem[Le~Bars(2018)]{le_bars_uncertainty_2018}
{\sc \au{Le~Bars, D.}} \yr{2018}  \at{Uncertainty in {Sea} {Level} {Rise} {Projections} {Due} to the {Dependence} {Between} {Contributors}}.  \jt{Earth's Future}  \bvol{6}~(9),  \pg{1275--1291}.

\bibitem[Malyarenko {\em et~al.\/}(2020)Malyarenko, Wells, Langhorne, Robinson, Williams \& Nicholls]{malyarenko_synthesis_2020}
{\sc \au{Malyarenko, A.}, \au{Wells, A.~J.}, \au{Langhorne, P.~J.}, \au{Robinson, N.~J.}, \au{Williams, M. J.~M.} \& \au{Nicholls, K.~W.}} \yr{2020}  \at{A synthesis of thermodynamic ablation at ice{\textendash}ocean interfaces from theory, observations and models}.  \jt{Ocean Model.}  \bvol{154},  \pg{101692}.

\bibitem[McCutchan {\em et~al.\/}(2024)McCutchan, Meyer \& Johnson]{mccutchan_enhancement_2024}
{\sc \au{McCutchan, Aubrey~L.}, \au{Meyer, Colin~R.} \& \au{Johnson, Blair~A.}} \yr{2024}  \at{Enhancement of ice melting in isotropic turbulence}.  \jt{Phys. Rev. Fluids}  \bvol{9}~(7),  \pg{074601}.

\bibitem[Meroni {\em et~al.\/}(2019)Meroni, McConnochie, Cenedese, Sutherland \& Snow]{meroni_nonlinear_2019}
{\sc \au{Meroni, Agostino~N.}, \au{McConnochie, Craig~D.}, \au{Cenedese, Claudia}, \au{Sutherland, Bruce} \& \au{Snow, Kate}} \yr{2019}  \at{Nonlinear influence of the {Earth}{\textquoteright}s rotation on iceberg melting}.  \jt{J. Fluid Mech.}  \bvol{858},  \pg{832--851}.

\bibitem[Millero \& Huang(2009)]{millero_density_2009}
{\sc \au{Millero, F.~J.} \& \au{Huang, F.}} \yr{2009}  \at{The density of seawater as a function of salinity (5 to 70 g/kg) and temperature (273.15 to 363.15 {K})}.  \jt{Ocean Sci.}  \bvol{5}~(2),  \pg{91--100}.

\bibitem[Orlowski(2012)]{orlowski_chasing_2012}
{\sc \au{Orlowski, J.}} \yr{2012} Chasing {Ice}.

\bibitem[Robel {\em et~al.\/}(2019)Robel, Seroussi \& Roe]{robel_marine_2019}
{\sc \au{Robel, A.~A.}, \au{Seroussi, H.} \& \au{Roe, G.~H.}} \yr{2019}  \at{Marine ice sheet instability amplifies and skews uncertainty in projections of future sea-level rise}.  \jt{Proc. Natl. Acad. Sci.}  \bvol{116}~(30),  \pg{14887--14892}.

\bibitem[Rothrock(1975)]{rothrock_mechanical_1975}
{\sc \au{Rothrock, D.~A.}} \yr{1975}  \at{The {Mechanical} {Behavior} of {Pack} {Ice}}.  \jt{Annu. Rev. Earth Planet. Sci.}  \bvol{3}~(1),  \pg{317--342}.

\bibitem[Rubinstein(1971)]{rubinstein_stefan_1971}
{\sc \au{Rubinstein, L.~I.}} \yr{1971} {\em The {Stefan} {Problem}\/}, ,  \vol{vol.~8}.  \publ{Providence: American Mathematical Soc.}

\bibitem[Russell-Head(1980)]{russell-head_melting_1980}
{\sc \au{Russell-Head, D.~S.}} \yr{1980}  \at{The {Melting} of {Free}-{Drifting} {Icebergs}}.  \jt{Ann. Glaciol.}  \bvol{1},  \pg{119--122}.

\bibitem[Schellenberg {\em et~al.\/}(2023)Schellenberg, Newton \& Hunt]{schellenberg_rotation_2023}
{\sc \au{Schellenberg, L.~M.}, \au{Newton, T.~J.} \& \au{Hunt, G.~R.}} \yr{2023}  \at{On the rotation of melting ice disks}.  \jt{Environ. Fluid Mech.}  \bvol{23}~(2),  \pg{465--488}.

\bibitem[Schloesser {\em et~al.\/}(2019)Schloesser, Friedrich, Timmermann, DeConto \& Pollard]{schloesser_antarctic_2019}
{\sc \au{Schloesser, F.}, \au{Friedrich, T.}, \au{Timmermann, A.}, \au{DeConto, R.~M.} \& \au{Pollard, D.}} \yr{2019}  \at{Antarctic iceberg impacts on future {Southern} {Hemisphere} climate}.  \jt{Nat. Clim. Chang.}  \bvol{9}~(9),  \pg{672--677}.

\bibitem[Silva {\em et~al.\/}(2006)Silva, Bigg \& Nicholls]{silva_contribution_2006}
{\sc \au{Silva, T. a.~M.}, \au{Bigg, G.~R.} \& \au{Nicholls, K.~W.}} \yr{2006}  \at{Contribution of giant icebergs to the {Southern} {Ocean} freshwater flux}.  \jt{J. Geophys. Res.: Oceans}  \bvol{111}~(C3).

\bibitem[Straneo \& Cenedese(2015)]{straneo_dynamics_2015}
{\sc \au{Straneo, F.} \& \au{Cenedese, C.}} \yr{2015}  \at{The {Dynamics} of {Greenland}'s {Glacial} {Fjords} and {Their} {Role} in {Climate}}.  \jt{Ann. Rev. Mar. Sci.}  \bvol{7}~(Volume 7, 2015),  \pg{89--112}, publisher: Annual Reviews.

\bibitem[Stroeve {\em et~al.\/}(2007)Stroeve, Holland, Meier, Scambos \& Serreze]{stroeve_arctic_2007}
{\sc \au{Stroeve, J.}, \au{Holland, M.~M.}, \au{Meier, W.}, \au{Scambos, T.} \& \au{Serreze, M.}} \yr{2007}  \at{Arctic sea ice decline: {Faster} than forecast}.  \jt{Geophys. Res. Lett.}  \bvol{34}~(9), \_eprint: https://onlinelibrary.wiley.com/doi/pdf/10.1029/2007GL029703.

\bibitem[Waasdorp {\em et~al.\/}(2024)Waasdorp, Bogaard, Wijngaarden \& Huisman]{waasdorp_melting_2024}
{\sc \au{Waasdorp, Pim}, \au{Bogaard, Aron van~den}, \au{Wijngaarden, Leen~van} \& \au{Huisman, Sander~G.}} \yr{2024}  \at{Melting of olive oil in immiscible surroundings: experiments and theory}.  \jt{Journal of Fluid Mechanics}  \bvol{998},  \pg{A18}.

\bibitem[Wells \& Worster(2008)]{wells_geophysical-scale_2008}
{\sc \au{Wells, A.~J.} \& \au{Worster, M.~G.}} \yr{2008}  \at{A geophysical-scale model of vertical natural convection boundary layers}.  \jt{J. Fluid Mech.}  \bvol{609},  \pg{111--137}.

\bibitem[Wengrove {\em et~al.\/}(2023)Wengrove, Pettit, Nash, Jackson \& Skyllingstad]{wengrove_melting_2023}
{\sc \au{Wengrove, M.~E.}, \au{Pettit, E.~C.}, \au{Nash, J.~D.}, \au{Jackson, R.~H.} \& \au{Skyllingstad, E.~D.}} \yr{2023}  \at{Melting of glacier ice enhanced by bursting air bubbles}.  \jt{Nat. Geosci.}  \bvol{16}~(10),  \pg{871--876}.

\bibitem[Xu {\em et~al.\/}(2024)Xu, Bootsma, Verzicco, Lohse \& Huisman]{xu_buoyancy-driven_2024}
{\sc \au{Xu, Dehao}, \au{Bootsma, Simen~T.}, \au{Verzicco, Roberto}, \au{Lohse, Detlef} \& \au{Huisman, Sander~G.}} \yr{2024} Buoyancy-driven flow regimes for a melting vertical ice cylinder in saline water. ArXiv:2410.22050 [physics].

\bibitem[Yamada {\em et~al.\/}(1997)Yamada, Fukusako, Kawanami \& Watanabe]{yamada_melting_1997}
{\sc \au{Yamada, M.}, \au{Fukusako, S.}, \au{Kawanami, T.} \& \au{Watanabe, C.}} \yr{1997}  \at{Melting heat transfer characteristics of a horizontal ice cylinder immersed in quiescent saline water}.  \jt{Int. J. Heat Mass Transf.}  \bvol{40}~(18),  \pg{4425--4435}.

\bibitem[Yankovsky \& Yashayaev(2014)]{yankovsky_surface_2014}
{\sc \au{Yankovsky, A.~E.} \& \au{Yashayaev, I.}} \yr{2014}  \at{Surface buoyant plumes from melting icebergs in the {Labrador} {Sea}}.  \jt{Deep-sea Res PT I}  \bvol{91},  \pg{1--9}.

\end{thebibliography}
\end{document}